\documentclass[sn-nature]{sn-jnl}

\usepackage{graphicx}%
\usepackage{siunitx,array,multirow}%
\usepackage{amsmath,amssymb,amsfonts}%
\usepackage{amsthm}%
\usepackage{mathrsfs}%
\usepackage{xcolor}%
\usepackage{textcomp}%
\usepackage{manyfoot}%
\usepackage{booktabs}%
\usepackage{algorithm}%
\usepackage{algorithmicx}%
\usepackage{listings}%

\usepackage{hyperref}
\usepackage{academicons}
\usepackage{subcaption}

\newcommand*{\figuretitle}[1]{%
    {
    \textbf{#1}
    \par\medskip}
}

\begin{document}

\title{A multi-cubic-kilometer neutrino telescope in the Western Pacific Ocean}


\author[1,2]{Z. P. Ye}\equalcont{These authors contributed equally to this work.}
\author[3]{F. Hu}\equalcont{These authors contributed equally to this work.}
\author[1,2]{W. Tian}\equalcont{These authors contributed equally to this work.}
\author[2]{Q. C. Chang}
\author[1]{Y. L. Chang}
\author[4,5]{\\Z. S. Cheng}
\author[2,6]{J. Gao}
\author[4]{T. Ge}
\author[7]{G. H. Gong}
\author[2]{J. Guo}
\author[4,5]{\\X. X. Guo}
\author[1]{X. G. He}
\author[1]{J. T. Huang}
\author[8,9]{K. Jiang}
\author[10]{\\P. K. Jiang}
\author[2,1]{Y. P. Jing}
\author[11]{H. L. Li}
\author[2]{J. L. Li}
\author[2]{L. Li}
\author[1]{\\W. L. Li}
\author[3,12]{Z. Li}
\author[4,5]{N. Y. Liao}
\author[8,9]{Q. Lin}
\author[13]{J. Lin}
\author[10]{F. Liu}
\author[2,1,14]{\\J. L. Liu}
\author[4]{X. H. Liu}
\author[8,9]{P. Miao}
\author[2]{C. Mo}
\author[1]{I. Morton-Blake}
\author[4]{\\T. Peng}
\author[2]{Z. Y. Sun}
\author[2]{J. N. Tang}
\author[8,9]{Z. B. Tang}
\author[15,16]{C. H. Tao}
\author[4,5]{\\X. L. Tian}
\author[2]{M. X. Wang}
\author[8,9]{Y. Wang}
\author[17]{Y. Wang}
\author[4,5]{H. D. Wei}
\author[2]{\\Z. Y. Wei}
\author[2]{W. H. Wu}
\author[1]{S. S. Xian}
\author[18,19,1,20]{D. Xiang}
\author*[1,2]{D. L. Xu}\email{donglianxu@sjtu.edu.cn}
\author[2]{\\Q. Xue}
\author[2]{J. H. Yang}
\author[4,5]{J. M. Yang}
\author[17]{W. B. Yu}
\author[15]{C. Zeng}
\author[1]{\\F. Y. D. Zhang}
\author[2]{T. Zhang}
\author[4,5]{X. T. Zhang}
\author[1]{Y. Y. Zhang}
\author[2]{\\W. Zhi}
\author[15]{Y. S. Zhong}
\author[15]{M. Zhou}
\author[21]{X. H. Zhu}
\author[4]{G. J. Zhuang} 


\affil[]{(\textbf{\large TRIDENT Collaboration)} \vspace{3pt}}

\affil[1]{\orgdiv{Tsung-Dao Lee Institute}, \orgname{Shanghai Jiao Tong University}, \orgaddress{\city{Shanghai}, \postcode{200240}, \country{China}}}
\affil[2]{\orgdiv{School of Physics and Astronomy, Shanghai Jiao Tong University, MOE Key Laboratory for Particle Astrophysics and Cosmology, Shanghai Key Laboratory for Particle Physics and Cosmology}, \orgaddress{\city{Shanghai}, \postcode{200240}, \country{China}}}
\affil[3]{\orgdiv{Department of Astronomy, School of Physics}, \orgname{Peking University}, \orgaddress{\city{Being}, \postcode{100871}, \country{China}}}
\affil[4]{\orgdiv{State Key Laboratory of Ocean Engineering, School of Naval Architecture Ocean and Civil Engineering}, \orgname{Shanghai Jiao Tong University}, \orgaddress{\city{Shanghai}, \postcode{200240}, \country{China}}}
\affil[5]{Shanghai Jiao Tong University Sanya Yazhou Bay Institute of Deep Sea Technology, Sanya, 572024, China}
\affil[6]{\orgdiv{Center for High Energy Physics}, \orgname{Peking University}, \orgaddress{\city{Beijing}, \postcode{100871}, \country{China}}}
\affil[7]{\orgdiv{Department of Engineering Physics}, \orgname{Tsinghua University}, \orgaddress{\city{Beijing}, \postcode{100084}, \country{China}}}
\affil[8]{\orgdiv{State Key Laboratory of Particle Detection and Electronics}, \orgname{University of Science and Technology of China}, \orgaddress{\city{Hefei}, \postcode{230026}, \country{China}}}
\affil[9]{\orgdiv{Department of Modern Physics}, \orgname{University of Science and Technology of China}, \orgaddress{\city{Hefei}, \postcode{230026}, \country{China}}}
\affil[10]{\orgdiv{Shanghai Key Lab of Electrical Insulation and Thermal Aging, School of Chemistry and Chemical Engineering}, \orgname{Shanghai Jiao Tong University}, \orgaddress{\city{Shanghai}, \postcode{200240}, \country{China}}}
\affil[11]{\orgdiv{Key Laboratory of Marine Ecosystem Dynamics}, \orgname{Second Institute of Oceanography, MNR}, \orgaddress{\city{Hangzhou}, \postcode{310012}, \country{China}}}
\affil[12]{\orgdiv{Kavli Institute for Astronomy and Astrophysics}, \orgname{Peking University}, \orgaddress{\city{Bejing}, \postcode{100871}, \country{China}}}
\affil[13]{\orgdiv{Center for High Performance Computing}, \orgname{Shanghai Jiao Tong University}, \orgaddress{\city{Shanghai}, \postcode{200240}, \country{China}}}
\affil[14]{\orgdiv{Shanghai Jiao Tong University Sichuan Research Institute}, \orgaddress{\city{Chengdu}, \postcode{610213}, \country{China}}}
\affil[15]{\orgdiv{School of Oceanography}, \orgname{Shanghai Jiao Tong University}, \orgaddress{\city{Shanghai}, \postcode{200030}, \country{China}}}
\affil[16]{\orgdiv{Key Laboratory of Submarine Geosciences}, \orgname{Second Institute of Oceanography, MNR}, \orgaddress{\city{Hangzhou}, \postcode{310012}, \country{China}}}
\affil[17]{\orgdiv{School of Electronic Information and Electrical Engineering}, \orgname{Shanghai Jiao Tong University}, \orgaddress{\city{Shanghai}, \postcode{200240}, \country{China}}}
\affil[18]{\orgdiv{MOE Key Laboratory for Laser Plasmas, School of Physics and Astronomy}, \orgname{Shanghai Jiao Tong University}, \orgaddress{\city{Shanghai}, \postcode{200240}, \country{China}}}
\affil[19]{\orgdiv{Collaborative Innovation Center of IFSA (CICIFSA)}, \orgname{Shanghai Jiao Tong University}, \orgaddress{\city{Shanghai}, \postcode{200240}, \country{China}}}
\affil[20]{\orgdiv{Zhangjiang Institute for Advanced Study}, \orgname{Shanghai Jiao Tong University}, \orgaddress{\city{Shanghai}, \postcode{200240}, \country{China}}}
\affil[21]{\orgdiv{State Key Laboratory of Satellite Ocean Environment Dynamics}, \orgname{Second Institute of Oceanography, MNR}, \orgaddress{\city{Hangzhou}, \postcode{310012}, \country{China}}}

\abstract{Next-generation neutrino telescopes with significantly improved sensitivity are required to pinpoint the sources of the diffuse astrophysical neutrino flux detected by IceCube and uncover the century-old puzzle of cosmic ray origins. A detector near the equator will provide a unique viewpoint of the neutrino sky, complementing IceCube and other neutrino telescopes in the Northern Hemisphere. Here we present results from an expedition to the north-eastern region of the South China Sea, in the western Pacific Ocean. A favorable neutrino telescope site was found on an abyssal plain at a depth of $\sim$ 3.5~km. At depths below 3 km, the sea current speed, water absorption and scattering lengths for Cherenkov light, were measured to be $v_{\mathrm{c}}<$~10~cm/s, $\lambda_{\mathrm{abs} }\simeq$~27~m and $\lambda_{\mathrm{sca} }\simeq$~63~m, respectively.  Accounting for these measurements, we present the design and expected performance of a next-generation neutrino telescope, TRopIcal DEep-sea Neutrino Telescope (TRIDENT). With its advanced photon-detection technology and large dimensions, TRIDENT expects to observe the IceCube steady source candidate NGC 1068 with 5$\sigma$ significance within 1 year of operation. This level of sensitivity will open a new arena for diagnosing the origin of cosmic rays and probing fundamental physics over astronomical baselines.}

\maketitle

\section*{Introduction}


Cosmic rays from deep space constantly bombard the Earth’s atmosphere, producing copious amounts of GeV – TeV neutrinos via hadronic interactions. Similar processes yielding higher energy (TeV – PeV) neutrinos are expected when cosmic rays are accelerated and interact in violent astrophysical sources, such as in jets of active galactic nuclei (AGN) \cite{Murase:2022agn}. Ultra-high-energy cosmic rays (UHECRs) traversing the Universe and colliding with cosmic background photons, are predicted to generate ‘cosmogenic’ neutrinos (ranging from PeV to ZeV) \cite{Kotera_2010}. Detecting astrophysical neutrino sources will therefore be the key to deciphering the origin of the UHECRs. 

The weak interactions which make neutrino detection so difficult, also allow them to be used as a powerful tool. Neutrinos can escape from extremely dense environments, travelling astronomical distances without being deflected or absorbed. Pointing back directly to their sources, neutrinos are a unique messenger to trace the most extreme regions of the Universe. Furthermore, neutrinos oscillate as they propagate through spacetime, transforming among flavours $\nu_e$, $\nu_\mu$ and $\nu_\tau$, due to the quantum effect known as flavor-mass mixing \cite{gdp_manual}. Measuring neutrino oscillation over astronomical baselines allows us to probe for new physics beyond the Standard Model \cite{Arguelles_snowmass:2022}, and also providing new handles for tests on quantum gravity \cite{IceCube:QG}.

Neutrinos cannot be detected directly. These ‘ghostly’ particles are measured using extremely sensitive technologies, detecting the light produced by the charged particles generated in neutrino-matter interactions. In a general detector setup, large areas of photon-sensors continuously monitor a large body of target mass, e.g. pure water \cite{pure_water_2003superK}, liquid scintillator \cite{liquid_scintillator_JUNO:2015}, liquid argon \cite{liquid_argon_DUNE:2015}, to measure these rare and tiny energy depositions. Neutrino telescopes use massive volumes of natural sea/lake water or glacial ice to observe the low rate of interacting high energy astrophysical neutrinos.

Theoretical calculations in 1998 suggested that a cubic-kilometer detector would be sufficiently sensitive to the high energy neutrino flux from AGN jets or gamma-ray bursts \cite{Waxman_Bahcall_bound}.
The IceCube Neutrino Observatory was the first experiment to build a telescope of this scale, instrumenting the deep glacial ice at the South Pole. IceCube made major breakthroughs over its lifetime, discovering a diffuse extraterrestrial neutrino flux in 2013 \cite{IceCube:2013low} and presenting compelling evidence for neutrino emission from a flaring blazar \cite{IceCube_TXS_flares:2018, IceCube_TXS_MM:2018} in 2017 and a Seyfert galaxy NGC 1068 in 2022 \cite{IceCube_NGC1068:2022}. Dedicated IceCube analyses, along with measurements made by the Astronomy with a Neutrino Telescope and Abyss environmental
RESearch (ANTARES) project in the Mediterranean Sea14 \cite{ANTARES:2011hfw}, have been carried out to resolve the origins of the diffuse cosmic neutrino flux. A wide range of hypotheses have been considered, including: all-sky spatial clustering searches \cite{IceCube_10yr_points_source:2020}, transient searches \cite{IceCube_GRB:2022, IceCube_FRB:2017, IceCube_FRB:2019} and AGN catalog stacking searches \cite{IceCube_AGN_jet:2016, IceCube_AGN_core:2021}, all yielding inconclusive results to date. This suggests multiple weaker sources \cite{Murase:2015xka} may contribute to the diffuse flux, such as star burst galaxies or AGNs \cite{starburst_galaxies:Ambrosone2021brr}, which would require better than $0.1^\circ$ pointing resolution to resolve \cite{Fang:2016hop}.

Several telescopes such as Cubic Kilometre Neutrino Telescope (KM3NeT) in the Mediterranean Sea \cite{KM3Net_letter_of_intent}, Baikal Gigaton Volume
Detector (Baikal-GVD) in Lake Baikal \cite{Baikal_GVD} and the newly proposed Pacific Ocean Neutrino Experiment (P-ONE) in the East Pacific \cite{P-ONE_nature}, are currently under development. 
Their northern locations will complement IceCube at the South Pole, offering full coverage of the teraelectronvolt-to-petaelectronvolt neutrino sky. Light propagating in the South Pole glacial ice generally experiences long absorption lengths and short scattering lengths. Conversely, deep-sea or lake water has longer scattering lengths but shorter absorption lengths. This reduced light scattering in water allows for substantial pointing resolution improvement in both the track and cascade channels, where the latter channel has been proven to have significantly lower contamination from atmospheric muons \cite{KM3NeT_reconstruction:2017,IceCube_HESE_7p5yr:2020}.

This work outlines our plan to construct a next-generation neutrino telescope in the South China Sea. The TRopical DEep-sea Neutrino Telescope (TRIDENT), nicknamed "Hai-Ling" in Chinese ("Ocean Bell") ({\url{https://trident.sjtu.edu.cn/en}}) aims to rapidly discover multiple high-energy astrophysical neutrino sources and significantly boost the measurement of cosmic neutrino events of all flavors. 
To achieve this goal, TRIDENT will instrument a massive volume of seawater and employ precise photon timing measurement to optimise its neutrino pointing resolution. Due to the Earth's rotation and TRIDENT's position near the equator, the detector's highest sensitivity band for up-going neutrinos will scan the entire sky, providing substantial visibility to all potential neutrino sources.

\section*{Results}\label{sec2}

A suitable site for constructing a deep-sea neutrino telescope demands multiple conditions. The depth should be large enough, e.g. $\gtrsim3~\mathrm{km}$, to effectively shield cosmic ray backgrounds and minimize the influence of biological activities. Experiences from the pioneering Deep Underwater Muon And Neutrino Detector (DUMAND) project ({\url{https://www.phys.hawaii.edu/~dumand/dumacomp.html}}) suggest that, a large and flat area such as an abyssal plain is preferred, and it should keep away from high rises or deep trenches to avoid complex current fields. The ocean floor should be flat and possess sufficient bearing strength to support the mounting of the equipment. On the basis of the successful operation of ANTARES for the past decade, a deep-sea neutrino telescope could safely operate under a current strength of less than $\sim20~\mathrm{cm/s}$ \cite{ANTARES:1999fhm}. Close proximity to a shore is required to ensure the infrastructure for power supply and data transmission via seafloor cable connections.  

On the basis of the above requirements, an area near 114.0$^\circ$E, 17.4$^\circ$N was selected as a suitable location to build a large scale deep-sea detector. The geographic information of the site is described in Methods and Extended Data Fig. \ref{fig:geographic_map}. Following the location selection, we carried out the TRIDENT pathfinder experiment (TRIDENT EXplorer, T-REX for short). With T-REX, we measured the optical properties of the sea water and also quantified the oceanographic conditions at the chosen site, including water current, temperature, salinity and radioactivity (see Methods as well as Extended Data Fig. \ref{fig:ocean_condition} and \ref{fig:k40_simulation}).

\subsection*{Optical properties of the deep-sea water}

Neutrino telescopes observe neutrino interactions by detecting Cherenkov photons generated in the medium. By measuring the number of these photons and their arrival times, information about the neutrino involved in the interaction can be reconstructed. To efficiently detect this light, excellent optical clarity is an important requirement in site selection.

The propagation of Cherenkov photons is predominantly affected by absorption and scattering. Absorption converts the photon energy into atomic heat via photon-molecule interactions \cite{mobley1994light}, which reduces the total number of observable photons. Scattering, on the other hand, causes photons to change their direction of propagation. This leads to a blurring of arrival times for Cherenkov light arriving at photon detection units, thus degrading the angular resolution of the neutrino telescope.

In sea water, scattering is dominated by two elastic processes, Rayleigh and Mie scattering, which can be quantified by their mean free path, denoted as $\lambda_\mathrm{Ray}$ and $\lambda_\mathrm{Mie}$, respectively. Mie scattering typically results in a small deflection angle and requires an additional parameter, $\cos\theta_{\mathrm{Mie}}$, to represent the mean scattering angle. The overall scattering effect, described by $\lambda_\mathrm{sca}$, can thus be expressed as: $1/\lambda_{\mathrm{sca}} = 1/\lambda_{\mathrm{Ray}} + 1/\lambda_{\mathrm{Mie}}$. Meanwhile, the absorption effect can be quantified using $\lambda_\mathrm{abs}$. As both absorption and scattering can occur during photon propagation, the concept of an attenuation length, $\lambda_{\mathrm{att}}$, is introduced to depict the exponential reduction in the intensity of a light beam within the medium, and can be formulated as: $1/\lambda_{\mathrm{att}} = 1/\lambda_{\mathrm{abs}} + 1/\lambda_{\mathrm{sca}}$.

When dealing with a spherical isotropic light source, however, it becomes more convenient to measure an effective attenuation length, $\lambda_{\mathrm{eff,att}}$, which approximately describes the decrease in the total observable photons $\propto e^{-D/\lambda_{\mathrm{eff,att}}}\cdot D^{-2}$ over a propagation distance $D$ \cite{ANTARES:2004kfl}. Notably, $\lambda_{\mathrm{eff,att}}$ differs from the canonical attenuation length $\lambda_{\mathrm{att}}$ since it also encompasses scattered photons in the observed light.

To decode all these optical parameters, precise in-situ measurements were conducted with T-REX, as shown in Extended Data Fig. \ref{fig:trex_apparatus}. 
The core detection unit consists of three modules. At the middle is a light emitter module equipped with light-emitting diodes (LEDs) of three wavelengths, which can emit photons isotropically with two modes: pulsing mode and steady mode. 
There are two light receiver modules located at $41.8~\mathrm{m}$ and $21.7~\mathrm{m}$ vertically above and below the light emitter, respectively, performing a near-far measurement. Both modules are equipped with two independent and complementary measurement systems, a photomultiplier tube (PMT) system and a camera system. 
The former primarily records PMT waveforms to extract the timing information of the detected photons emitted by pulsing LEDs, while the latter records images of the steady light emitter to measure the angular distribution of the radiance (Methods and Extended Data Fig. \ref{fig:pmt_fitting} and \ref{fig:3420m_images}).

Table \ref{tab:results} summarizes the measured canonical optical parameters using both the PMT and camera systems in the blue waveband, the optimal waveband for observing Cherenkov photons in water. The two systems work independently and obtain consistent results using different measurement mechanisms. All of the data processing and analysis pipelines are presented in Methods in detail. In addition, measurement results at three different wavelengths, at various depths, are listed in Extended Data Table \ref{tab:pmt_extend_results} and \ref{tab:camera_extend_results}.

\begin{table}[ht!]
\figuretitle{Optical parameters measured in the blue waveband}
\begin{subtable}[t]{\textwidth}
    \caption{Optical parameters measured by the PMT system at 450 nm.}
    \begin{tabular*}{\textwidth}{@{\extracolsep{\fill}}lrrrrr}
    \toprule
    Method & $\lambda_{\text{abs}}$ [m] & $\lambda_{\text{ray}}$ [m] & $\lambda_{\text{mie}} $[m] & $\cos \theta_{\text{mie}}$ & $\lambda_{\text{att}}$ [m] \\
    \midrule
    $\chi^2$ fitting & $ 27.4 ^{+1.1}_{-0.9}$ & $ 200 ^{+13}_{-10}$ & $ 84^{+12}_{-8}$ & $0.97 ^{+0.02}_{-0.02}$ & $18.7^{+3.0}_{-2.1}$ \\ 
    MCMC & $26.4^{+1.2}_{-1.0}$ & $203^{+15}_{-11}$ & $64^{+12}_{-14}$ & $0.97^{+0.01}_{-0.01}$ & $17.2^{+0.8}_{-1.3}$  \\
    \bottomrule
    \end{tabular*}
\end{subtable}
\vspace*{5mm}
\begin{subtable}[t]{\textwidth}
    \caption{Optical parameters measured by the camera system at 460 nm. }
    \begin{tabular*}{\textwidth}{@{\extracolsep{\fill}}lccc}
    \toprule
    Method & $\lambda_{\text{abs}}$ [m] & $\lambda_{\text{sca}}$ [m] & $\lambda_{\text{att}}$ [m] \\
    \midrule
    $\chi^2$ fitting & $26.5\pm0.5$ & $62.9\pm3.7$ & $18.7\pm0.2$ \\
      $I_{\text{center}}$ & $-$ & $-$ & $19.3 \pm 1.3$  \\
    \bottomrule
    \end{tabular*}
\end{subtable}
\caption{Measurement results of optical properties for seawater at the depth of $3420~\mathrm{m}$ for both the PMT ($450~\mathrm{nm}$) and camera ($460~\mathrm{nm}$) systems. The data collection for the PMT and camera system lasted $\sim 50$ min and $\sim 8$ min, respectively. The table displays data along with error bars that indicate mean values and 68\% confidence intervals. These error bars consider both statistical and systematic uncertainties.}
\label{tab:results}
\end{table}

\begin{figure}[htbp]%
    \centering
    \figuretitle{Measured (effective) attenuation length at different neutrino telescope sites}
    \includegraphics[width=0.98\textwidth]{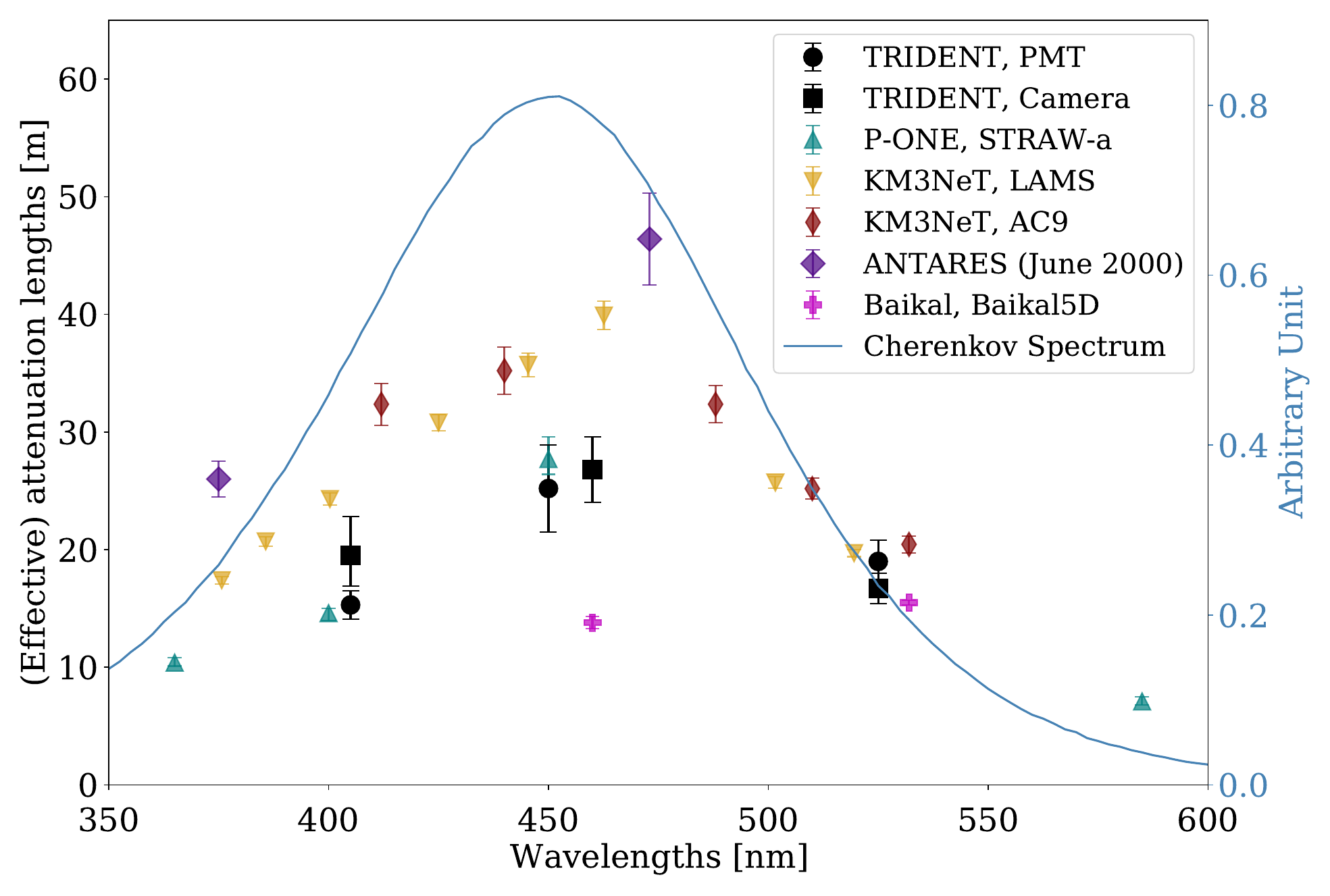}
    \caption{Effective attenuation lengths measured by two independent optical systems, the PMT (black circle) and camera (black square) in T-REX for three wavelengths (405~nm, 450/460~nm, 525~nm). Error bars reflect both statistical and systematic uncertainties. Measurements from KM3NeT \cite{KM3NeT:2011ioj, Riccobene:2006qy}, P-ONE \cite{Bailly:2021dxn}, ANTARES \cite{ANTARES:2004kfl}, and Baikal-GVD \cite{baikal_neutrino2022} are shown for comparison. Also shown is the observed Cherenkov spectrum from simulation, in which the optical properties measured by T-REX are implemented.}
    \label{fig:attenuation}
\end{figure}

\begin{table}[htpb]
\figuretitle{Effective attenuation lengths measured at various wavelengths}
\centering
    \begin{tabular*}{\textwidth}{@{\extracolsep{\fill}}lcccc}
      Wavelengths [nm] & 405 nm & 450 nm & 460 nm & 525 nm  \\   
     \midrule
      PMT & $15.3\pm 1.2$ & $25.2\pm 3.7$ & $-$ & $19.0\pm1.8$ \\ 
      Camera & $19.5^{+3.3}_{-2.6}$ & $-$ & $26.8\pm 2.8$ & $16.7\pm 1.3$  \\
      \bottomrule
      \end{tabular*}
      \caption{Effective attenuation lengths measured by the PMT and camera systems at different wavelengths. The table displays data along with error bars that indicate mean values and 68\% confidence intervals. These error bars consider both statistical and systematic uncertainties.}
      \label{tab:eff_att}
\end{table}  

Fig. \ref{fig:attenuation} summarizes the measurement results of optical property at TRIDENT's site and other water-based neutrino telescopes' sites.
To compare with other similar measurements, we conducted another set of analyses to obtain $\lambda_{\text{eff,att}}$, as listed in Table \ref{tab:eff_att}, as definitions of the attenuation length in other experiments differ slightly. 
The results from Long Arm Marine Spectrophotometer (LAMS) \cite{KM3NeT:2011ioj}, ANTARES \cite{ANTARES:2004kfl} and STRings for Absorption length in Water (STRAW-a) \cite{Bailly:2021dxn} are effective attenuation lengths, which contain different proportions of scattered photons in their selected data acquisition time windows. The results from Baikal-5D \cite{baikal_neutrino2022} and AC9 \cite{Riccobene:2006qy}, however, made measurements of canonical attenuation lengths using specialized laser devices. 

The measured optical properties and water current speeds are promising for operating a large scale neutrino telescope at the selected site. 
T-REX's camera system demonstrated its application as a fast, in-situ calibration system, which is particularly important for precise angular reconstruction in underwater telescopes with dynamic environments. 
Additionally, T-REX has been a valuable tool for testing some of TRIDENT's electronic systems, such as time synchronization technologies and optical fibers for data transmission.

\subsection*{Design of TRIDENT}

\begin{figure}[!htb]%
    \centering
    \figuretitle{Geometrical layout of the TRIDENT array}
    \includegraphics[width=0.85\textwidth]{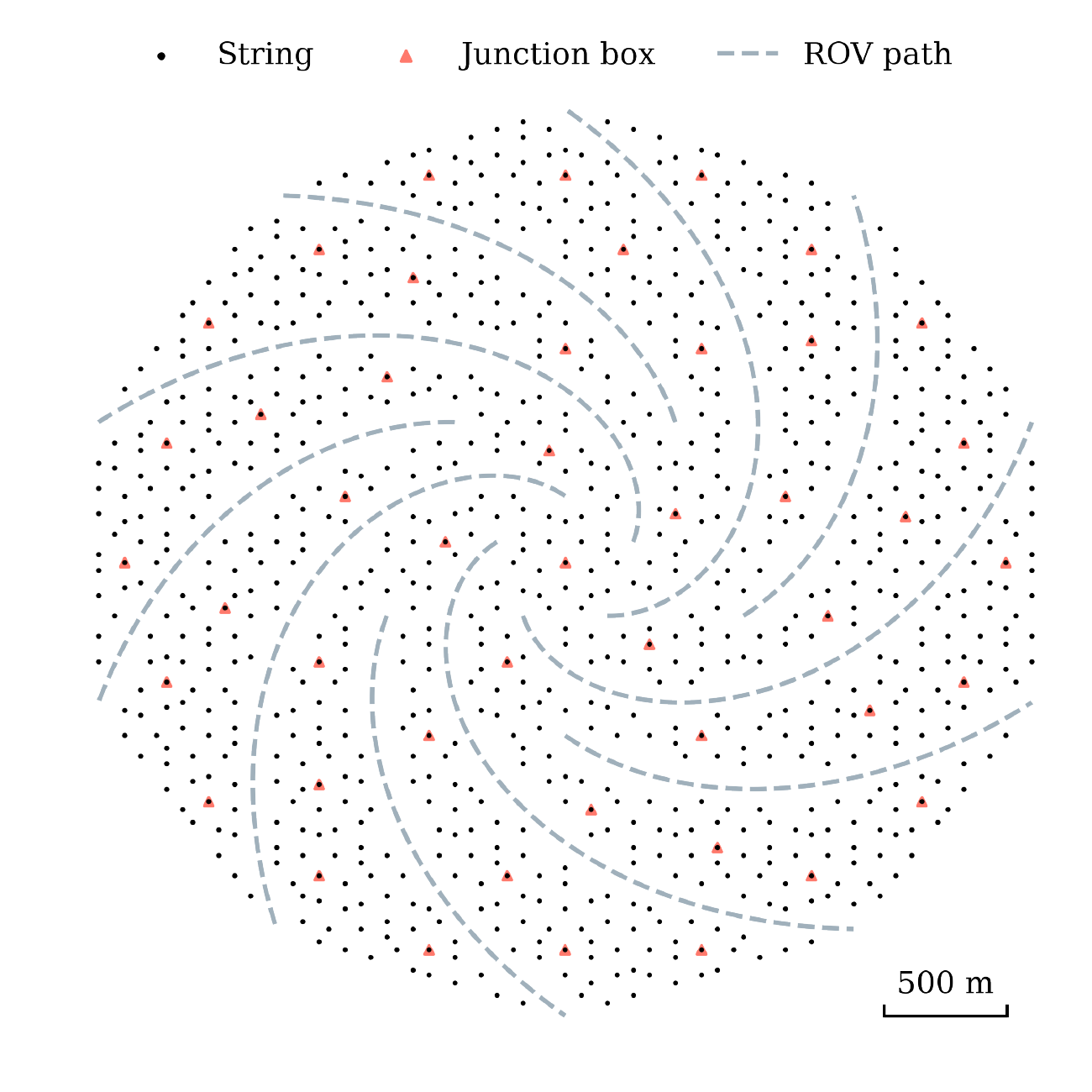}
    \caption{Geometry layout of the TRIDENT array. The pattern follows a Penrose tiling distribution. Each black dot represents a string of length $\sim0.7~\mathrm{km}$, while the dashed lines mark the paths for underwater maintenance robots.}
    \label{fig:geo-layout}
\end{figure}

TRIDENT will be optimized to pinpoint astrophysical neutrino sources from the isotropic diffuse flux discovered by IceCube. 
The long scattering lengths in deep-sea water allow the Cherenkov photons from a neutrino interaction vertex to propagate in long straight paths to the many optical sensors throughout the detector.
Precisely measuring the arrival times of these direct photons strongly improves the angular resolution of track-like events due to $\nu_\mu$ (and a fraction of $\nu_\tau$) charged current interactions, which neutrino telescopes rely primarily on for pointing \cite{Wiebusch:2003}.
TRIDENT aims to achieve this with the help of modern silicon photon multipliers (SiPMs) that can respond to photon hits within tens of picoseconds \cite{SiPM_tech}, time digital converters that are capable of digitizing the sharp rising edge of a SiPM waveform \cite{TDC_tech} and the White Rabbit system that can provide precise global time stamps \cite{Wlostowski:2022fau}. 
With these state-of-the-art technologies, TRIDENT will build hybrid digital optical modules with both PMTs and SiPMs, called hDOMs \cite{Hu:2021jjt}, yielding excellent light collection and timing resolution that are capable of accurately measuring the arrival time of unscattered photons. The advantages of using multiple small PMTs have been demonstrated by KM3NeT's multi-PMT Digital Optical Module (mDOM) system \cite{KM3NeT:2022pnv}. Compared to IceCube's single large PMT DOM, multiple small PMTs allow for an increased photocathode coverage, strong sensitivity to the incident photon direction, finer timing resolution, along with the capability of coincidence triggering on a single DOM. In an effort to further improve angular resolution, the TRIDENT hDOM design adds SiPMs with excellent timing resolution, placed in the spaces between PMTs. The first-rate timing response and additional photocathode coverage of the SiPMs in TRIDENT’s hDOM design are expected to provide improvement in angular resolution compared to traditional PMT-only DOMs, boosting the detector's source searching ability. 

In seawater, the absorption length for Cherenkov photons is a key parameter to consider when designing the detector geometry. 
Fig. \ref{fig:geo-layout} shows the anticipated layout of the future telescope, guided by the presented optical property measurements. 
The detector contains 1211 strings, each containing 20 hDOMs separated vertically by $30~\mathrm{m}$, ranging from approximately 2800 m to 3400 m below sea level. This arrangement will result in a world-leading instrumented geometric volume of $\sim 7.5~\mathrm{km}^3$. 
The strings' pattern follows a Penrose tiling distribution with inter-string distances of $70~\mathrm{m}$ and $110~\mathrm{m}$, adopting the golden ratio \cite{tilings_and_patterns}. 
Preliminary simulation studies indicate that this uneven layout, compared to a regular distribution of strings, allows for an expanded geometry with a broader window of measurable neutrino energies. 
TRIDENT in this layout is expected to cover from sub-teraelectronvolt (TeV) to exaelectronvolt energies, optimizing the telescope's potential for neutrino astronomy \cite{Fang:2022trf}. 
Building multiple, separated clusters of strings helps to ease the difficulties faced in the construction and maintenance of large telescopes on the seafloor. 
TRIDENT instead leaves several spiral pathways, allowing underwater robots to access the innermost strings for maintenance. 
This unsegmented geometry aims to reduce the number of clipping edge events, which are more likely to occur in segmented geometries with wide empty regions between string clusters. 
The spiral shape of the pathways also helps to reduce the number of `corridor events', which describe undetected muons passing straight through parallel arrays of strings. 
Acoustic detectors will be installed on each string for high-precision position calibration. 
These detectors can also be placed sparsely in an array extended beyond the main detector volume, to detect cosmogenic neutrinos with energies well above exaelectronvolts \cite{ANTARES:2010, Neff_2013}.

\subsection*{Source sensitivity and discovery potentials}

\begin{figure}[!htbp] 
    \centering
    \figuretitle{Projected point source sensitivities and discovery potentials of TRIDENT}    
    \includegraphics[width=0.98\linewidth]
    {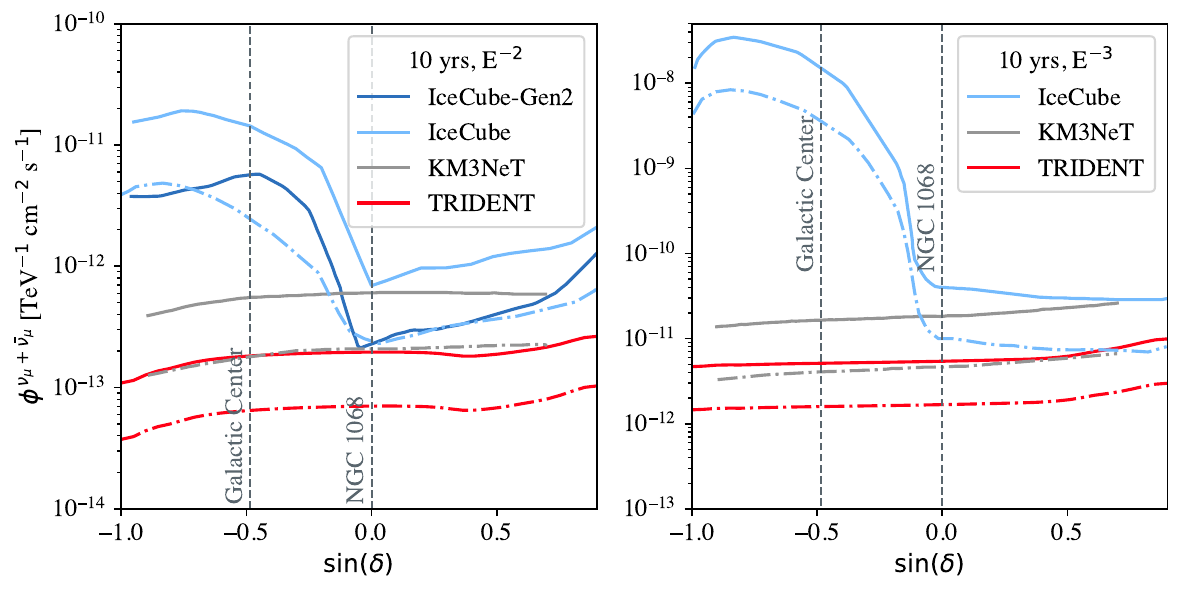}
    \label{fig:allsky_sensitivity}
    \caption{
    All sky point source 90\% confidence-level median sensitivity (dashed dot lines) and $5\sigma$ discovery potential (solid lines) of TRIDENT with 10 years of data taking. 
    The left panel corresponds to a source spectrum index of 2 (labelled $E^{-2}$) and minimum energy of $10~\mathrm{TeV}$, while the right assumes an index of 3 ($E^{-2}$) and minimum energy of $1~\mathrm{TeV}$. 
    The $x$ axis represents the sine declination ($\sin\delta$) and the $y$ axis is the neutrino flux ($\phi$). 
    KM3NeT, IceCube and IceCube-Gen2 sensitivities \cite{KM3NeT_sensitivity:2018, IceCube_10yr_points_source:2020, IceCube-Gen2:geometry} are also shown for comparison. IceCube, located at the South Pole, has increased sensitivity to the northern sky. For a source located in the southern sky with a spectral index of 3, TRIDENT will have 4 orders of magnitude improvement in sensitivity compared to IceCube. Similarly comparing to the future telescope KM3NeT located in the northern hemisphere, yields an improvement factor of approximately 5.}
    \label{fig:sensitivity_DisPot_all_sky}
\end{figure}

\begin{figure}[!htbp] 
    \centering
    \figuretitle{Projected discovery potentials of TRIDENT for potential neutrino sources}
    \includegraphics[width=0.65\textwidth]{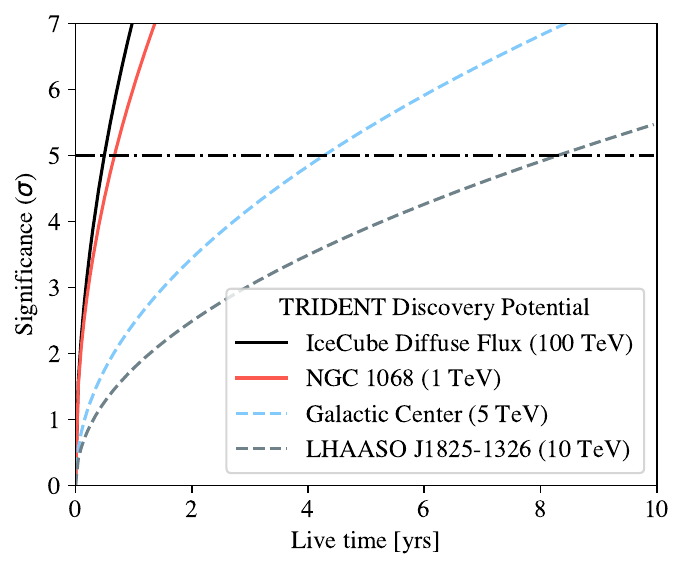}
    \caption{The source flux used in this figure: NGC 1068 flux from IceCube best fit result with spectrum index of 3.2 \cite{IceCube_NGC1068:2022}; diffuse neutrino flux from IceCube best fit result \cite{IceCube_diffuse_tracks:2022}; Galactic Center from conversion of High Energy
Stereoscopic System (HESS) gamma-ray observation to neutrino flux upper limit with gamma-ray cut-off energy at $100\,\mathrm{TeV}$ \cite{HESS_GC:2016, Celli:2016}; Large High Altitude Air Shower Observatory (LHAASO) J1825-1326 from conversion of LHAASO gamma-ray observation to neutrino flux upper limit with cut-off energy at $286\,\mathrm{TeV}$ \cite{LHAASO:2021gok, Huang:2021}. The horizontal black dashed line indicates a significance level of $5\sigma$. In the analysis, the diffuse astrophysical muon neutrinos \cite{IceCube_diffuse_tracks:2022} and atmospheric muon neutrinos \cite{IceCube_atmos_neutrino:2016} are considered as backgrounds. The minimum energies adopted for each source are shown in the legend.}
    \label{fig:sensitivity_DisPot_specific_sources}
\end{figure}

We conducted a performance study of TRIDENT using in-depth simulations, in particular measuring up-going (neutrinos with zenith angle greater than $80^\circ$ and travel long distances inside Earth) track events produced in $\nu_{\mu}$ charged-current interactions. 
At an energy of $\sim100~\mathrm{TeV}$, the angular resolution of TRIDENT is expected to reach 0.1$^{\circ}$ with an effective area of $5 \times 10^2 ~\mathrm{m^2}$, as shown in 
Extended Data Fig. \ref{fig:angular_resolution} and \ref{fig:effective_area}. 
TRIDENT intends to push the limits of neutrino telescope performance, reaching a new frontier of sensitivity in all-sky searches for astrophysical neutrino sources, as shown in Fig. \ref{fig:sensitivity_DisPot_all_sky}.

The most promising neutrino source candidates found by IceCube, NGC 1068 and TXS 0506+056 \cite{IceCube_NGC1068:2022, IceCube_TXS_flares:2018, IceCube_TXS_MM:2018}, will be visible to TRIDENT in the up-going neutrino mode, for $\sim$ 50\% of its operation time.
Assuming an IceCube best fit flux, TRIDENT is predicted to discover the steady source NGC 1068 within one year of operation, as shown in Fig. \ref{fig:sensitivity_DisPot_specific_sources}. 
For a transient neutrino burst similar to the TXS 0506+056 2014-2015 case, TRIDENT will detect it with a significance level over $10\sigma$.

\subsection*{Physics with all neutrino flavors}

Once astrophysical neutrino sources are identified, oscillation physics and searches for new physics will become feasible by measuring neutrino flavor ratios at fixed astronomical baselines.
In particular, we expect to significantly boost measurements of astrophysical tau neutrino events using modern long waveform readout electronics \cite{IceCube_double_waveform:2015prd, SCA_tech} and also identify astrophysical electron antineutrinos via the Glashow resonance channel \cite{Glashow:1960}.
It is a particularly exciting time because IceCube has seen the first evidence for both types of events \cite{juliana_tau, icrc_max, icrc_logan, IceCube_Glashow:2021}. Drastically boosting the statistics of these event types will produce a plethora of new physics opportunities.
Another aspect to improve is the discrimination efficiency between electromagnetic and hadronic showers from $\nu_e$ CC interactions and neutral-current interactions of all flavors by their distinct particle compositions \cite{neutron_echo}.
This will open unique windows via the weak sector to probe physics at an energy frontier out of reach by contemporary man-made accelerators \cite{Sun:2022lti, Zhou:2021xuh}.

\subsection*{Ocean engineering and timelines}
It is not a trivial task to construct and operate such a giant array of precision detectors in a highly dynamic water body. 
Each string is mounted to the seafloor and raised by a buoy tied to the other end. Care should be taken to achieve the correct buoyancy and cable strength, accounting for possible extreme conditions such as benthic storms in the abyss. 
We carried out the first batch of small-scale tests of the array-current response, using $1:25$ scale string toy models in a ship towing tank (\url{https://oe.sjtu.edu.cn/EN/}) located on the campus of Shanghai Jiao Tong University (Supplementary Video 1). For current speeds at $10~\mathrm{cm/s}$, horizontal string displacements were found to be less than $30~\mathrm{cm}$.
More dedicated testing will be employed to guide the mechanical design of the strings. 
Modules monitoring the sea currents will be installed among the telescope array to track the instantaneous dynamics of the environment, ensuring smooth operation. 

Following hDOM prototype development, a dedicated factory will be built for the mass production and testing of the hDOM and string systems in the port city of Sanya. From there they can be conveniently shipped out to site for deployment. A pilot project with 10 strings installed in the selected site for a technology demonstration is scheduled for $\sim$2026. Construction of the full array can begin following a successful demonstration, commencing measurements in its partially built configuration. The full telescope is envisioned to become live in the early 2030s.

\section*{Methods}

\subsection*{Geographic information}

Based on the existing geographic and oceanographic data \cite{site_info}, we identified a huge abyssal plain in the northern part of the South China Sea that can meet the critical conditions for building a neutrino telescope. A uniform $10 \times 10~\mathrm{km}$ area was investigated on this plain, measured at a depth of $3475 \pm 8~\mathrm{m}$, which can largely avoid biological activities and provides sufficient overburden to shield cosmic ray muons down to $\sim 10^{-8}$ cm$^{-2}$s$^{-1}$sr$^{-1}$ \cite{Bugaev:1998bi}. The seabed of this area is mainly covered by clay silt and the mean slope is only $0.01^{\circ}$. Such a flat seabed ensures a uniform current distribution. A long-term high-resolution simulation using the Regional Ocean Modeling Systems \cite{site_simulation} indicates that the average bottom current speed is only $6~\mathrm{cm/s}$ over the past 30 years and the maximum value is $\sim 26~\mathrm{cm/s}$. The selected site is 180 km away from the Yongxing Island, where power supply and data transmission over this distance is feasible (Extended Data Fig. \ref{fig:geographic_map}).

\subsection*{Oceanographic conditions and radioactivity}

The water current at different depths was measured at the chosen site using an LADCP (Lowered Acoustic Doppler Current Profiler) on 6 September, 2021, shown in the lower panel of Extended Data Fig. \ref{fig:ocean_condition}.
Below $\sim 2800 ~\mathrm{m}$, the water current speed is less than $10~\mathrm{cm/s}$. Its direction and gradient as a function of depth has a smooth profile. And the current is steady, changing slowly in time. Additionally, the temperature and salinity, measured by a CTD (Conductivity, Temperature and Pressure), are shown in the upper panel of Extended Data Fig. \ref{fig:ocean_condition}. Below $2500~\mathrm{m}$, the temperature becomes constant at $\sim 2^{\circ}\mathrm{C}$.

To precisely measure the radioactivity of the site, in-situ water was collected with the CTD and transported back to Shanghai through cold chain logistics. 
The radioactivity (predominantly $^{40}$K) of the water was then measured by the PandaX team using high purity germanium detector \cite{PandaX_radioactivity} in the China Jinping Underground Laboratory. 
The measured abundance of $^{40}\mathrm{K}$ is $10.78\pm0.21~\mathrm{Bq/kg}$, consistent with public data of ordinary seawater. 
A background analysis with \textsf{Geant4} \cite{GEANT4:2002} simulation indicates that this level of radioactivity corresponds to a trigger rate of $\sim 2~\mathrm{kHz}$ per single 3-inch PMT assumed with $29\%$ quantum efficiency at 450 nm, as shown in Extended Data Fig. \ref{fig:k40_simulation}, acceptable for operating both T-REX and TRIDENT. 
During the apparatus deployment and data taking periods, marine life occasionally showed up and was recorded by our live-cameras only above a depth of $1500~\mathrm{m}$. 
No activity was spotted during the 0.5 h camera data-taking period at depths below $3000~\mathrm{m}$.

\subsection*{Deployment of T-REX and data taking}

In order to safely deploy the long and delicate apparatus, T-REX, shown in Extended Data Fig. \ref{fig:trex_apparatus}, was first packed on the deck like a wire roller, then hoisted into the water. It then unfolded naturally under the action of its buoyancy and gravity. The ballast at the bottom weighs about $700~\mathrm{kg}$, and the connection cables between the detection modules are made of high-rigidity steel wires to ensure that the low-frequency disturbance of the research vessel will not excite the resonance of the system.

During the deployment process, planned tests were conducted by the camera system at the fixed depths of $1221~\mathrm{m}$ and $2042~\mathrm{m}$. 
Each test took $8~\mathrm{min}$ to record data. After reaching the target depth of $3420~\mathrm{m}$, the whole apparatus was suspended for $\sim$ 2 hours to conduct in-situ measurements. 
The data taking was then divided into two stages. The light emitter was first operated in the pulsing mode to trigger the PMT system, which lasted $\sim$ 1.5 h. 
For the wavelength of $450~\mathrm{nm}$, it took about $50~\mathrm{min}$ to collect more than $10^{7}$ photons. 
The data collection for the other two wavelengths, namely $405~\mathrm{nm}$ and $525~\mathrm{nm}$, lasted about $10~\mathrm{min}$ for quick measurements.  
In the second stage, the camera system recorded more than 3000 images in $\sim$ 0.5 h when the light emitter was switched to steady mode with wavelengths of $460~\mathrm{nm}$, $525~\mathrm{nm}$ and $405~\mathrm{nm}$ in sequence. 
After completing the measurements at the depth of $3420~\mathrm{m}$, the whole apparatus was retrieved for recycle. 

\subsection*{PMT data analysis} 

Three inch PMTs and pulsing LEDs are synchronically triggered by the White Rabbit system at a rate of $10~\mathrm{kHz}$. The spread of the LED pulses is 3 ns \cite{Li:2023wqk, Tang:2023jmn}. For non-scattered light, the PMT will observe the narrow pulses; for scattered light, the photons will arrive later at the PMT, forming a scattering tail in the photon arrival time distribution. A $1000~\mathrm{ns}$ data acquisition (DAQ) \cite{Wang:2023rvb} window is set for the PMTs to observe the light from each LED pulse.

PMT data analysis is done in two steps. First, the photon arrival time distribution for each PMT is obtained. Second, the distribution is fitted with light propagation models simulated by \textsf{Geant4} for various optical parameters \cite{Hu:2023ife}. The ratio of the number of photons detected by the top and bottom PMTs, after correcting the square of distance, shows the effect of absorption. The tail of photon arrival time distribution contains the information of scattering.

The photon arrival time distribution is obtained by getting the times of PMT signals and the number of photo-electrons ($N_\mathrm{pe}$) in each signal and then filling the signal times in a histogram with $N_\mathrm{pe}$ as the weight. PMT signals are found in the waveform if the voltage is higher than one-third of the single photo-electron amplitude. The waveform peak time is determined as the signal time. The waveform is integrated to get the charge, and dividing the charge by the PMT gain yields $N_\mathrm{pe}$ for each signal. Uncertainties in the signal time, charge integral and PMT gain are considered in the analysis. See Extended Data Fig. \ref{fig:pmt_fitting} for an example of the derived photon arrival time distributions.

The model used to fit the photon arrival time distribution is composed of (1) emission from the pulsing LEDs, (2) light propagation in the water and (3) detection by the PMTs. It can be written as: 

\begin{equation}
[ N_{\mathrm{emi}} \times f(t) ] \otimes [ \frac{A}{4 \pi D^2} \times P( t \,\vert\, \lambda_\mathrm{abs} , \lambda_\mathrm{Ray} , \lambda_\mathrm{Mie} , \left \langle \cos\theta_{\mathrm{Mie}} \right \rangle , n , D ) ] \otimes [ \eta \times g(t) ],
\end{equation}

\noindent where $N_{\mathrm{emi}}$ is the number of emitted photons, $f(t)$ is the LED pulse timing profile, $A$ is the effective detection area of the PMT, $D$ is the distance, $P( t \,\vert\, \lambda_{\mathrm{abs}} , \lambda_\mathrm{Ray} , \lambda_\mathrm{Mie} , \left \langle \cos\theta_{\mathrm{Mie}} \right \rangle , n, D )$ is the photon propagation function, $\eta$ and $g(t)$ are the PMT detection efficiency and time response, respectively. The convolution of the LED pulse profile and PMT time response $f(t) \otimes g(t)$ is measured in the lab. The final model is computed by convolving the simulated photon propagation function and calibrated LED and PMT time response. The blocking effect of cable is less than 0.1\% and is neglected in the analysis.

A $\chi^2$ fitting method is adopted to fit the photon arrival time distribution with the above model for a pair of PMTs from the top and bottom receiver modules:

\begin{equation}
{\chi}^2 = \sum_{i=1}^{N}{ \frac{ ( D_i - M_i - \sum_{k=1}^{K}{ c_k \cdot \beta_{ki} } )^2 }{ {\sigma_i}^2 } } + \sum_{k=1}^{K}{ {c_k}^2 } , 
\end{equation}

\noindent where $D_i$ is the number of photons in the $i^\mathrm{th}$ bin of the photon arrival time distribution and $M_i$ is the expected value by the model. The uncorrelated uncertainty $\sigma_i$ includes statistical fluctuation, electronic noise and uncertainty of LED pulse profile and PMT time response. 
Correlated uncertainties ($\beta_{ki}$) include fluctuations of LED brightness, distances, PMT gain, PMT detection efficiencies, and the binning effect caused by the $4\,\text{ns}$ ADC resolution.
Minimization of the ${\chi}^2$ will return the best-fit model and yield the measurement results and uncertainties for the physics parameters: $\lambda_\mathrm{abs}$, $\lambda_\mathrm{Ray}$, $\lambda_\mathrm{Mie}$, $\left \langle \cos\theta_{\mathrm{Mie}} \right \rangle$, and refraction index $n$.

For cross validation, a Markov Chain Monte Carlo (MCMC) technique is also used to obtain the best model using the \textsf{emcee} sampler \cite{foreman2013emcee}. The goal of MCMC is to approximate the posterior distribution of model parameters by random sampling in a probabilistic space. A multi-dimension linear interpolation is performed before the sampling since the model in our case is discrete. 

Both the $\chi^2$ fitting and MCMC methods follow the same analysis procedure described above but have minor differences in the detailed treatment of convolution. 
Despite this, the two methods yield consistent results. Three pairs of PMTs from the top and bottom light receivers are used to fit the optical parameters independently, yielding consistent results as well.

The effective attenuation length is derived by comparing the number of photons $N_\mathrm{hit}$ received by top and bottom PMTs. 
Here, $N_\mathrm{hit}$ is the integral of photon arrival time distribution over the DAQ window. Simulation studies show that $\sim25\%$ and $\sim45\%$ of the $N_\mathrm{hit}$ photons are scattered at least once, for the near and far PMT respectively, 
making the effective attenuation length deviate from the canonical attenuation length, as discussed before \cite{Hu:2023ife}.

\subsection*{Camera data analysis} 
It has been discussed extensively among the neutrino telescope community, including IceCube-Gen2, KM3NeT and P-ONE, that cameras can be an excellent tool for optical calibration and \textit{in-situ} environment monitoring. TRIDENT plans to use a dedicated camera system for fast optical calibration when it becomes operational. The prototype for this camera system, including both hardware design and data processing pipeline, have been fully tested during the T-REX explorer mission. The camera system of T-REX mainly consists of a 5-million-pixel monochromatic camera with a fixed-focus lens. It is controlled by a Raspberry-4Pi module which can transfer its real-time data back to the research vessel \cite{Wang:2023rvb}. Both top and bottom cameras (denoted as \text{CamA} and \text{CamB}) are calibrated to have a proper viewing angle of about $16^\circ$ in the seawater. During the data-taking process, the same series of exposure times of 0.02s, 0.05s, 0.07s, 0.11s and 0.2s were configured for all three wavelengths, allowing an adequately broad range to accommodate for the blind conditions in the deep water. The relatively short exposure times can reduce the potential motion blur caused by sea current perturbations affecting the whole apparatus, but are long enough to provide sufficient photon statistics as the camera system operates under the steady mode of the light emitter \cite{Li:2023wqk, Tang:2023jmn}. The key observable for the camera system is the angular distribution of the radiance, which is converted into gray values of pixels.

The first method for the camera system, called the $I_\mathrm{center}$ method, is used to quickly measure the attenuation length of the medium by comparing the gray values in the central region of light emitter images:
\begin{equation}
\lambda_{\text{att}} = -(D_{A}-D_{B}) /\ln \left(-\frac{I_{\text{A}}}{I_{\text{B}}}\times \frac{I_{0}'}{I_{0}}\right) .
\label{I_center}
\end{equation}
Here, $D_{A} - D_{B}$ is the relative distance between the two cameras. $I_{\mathrm{A}}$ and $ I_{\mathrm{B}}$ are the mean gray values in the center region of images from \text{CamA} and \text{CamB}, corresponding to the directly arriving light from the emitter. $I_{0}'/ I_{0}$ is the initial intensity ratio of both sides of the light emitter, which is well calibrated. The canonical attenuation length can be derived from such a setup due to the far distances between the cameras and light emitter, which makes the directly arriving light highly collimated. The open angle of the light emitter is $<1.1^{\circ}$ for \text{CamA} and $<0.6^{\circ}$ for \text{CamB}, thus both absorption and scattering will dissipate the radiance. A \textsf{Geant4} simulation study shows that there is a small contamination of scattered photons in the centroid pixel, but this contamination ratio is approximately equal in both the near and far cameras, causing an uncertainty of less than 4\% on the $I_\mathrm{center}$ method \cite{Hu:2023ife}.

For the image processing, we select images with suitable exposure time and gain, to ensure that all the pixels are within the range of linear response. We then find the centroid of each image and crop it to a unified size of $300\times300$ pixels, shown in Extended Data Fig. \ref{fig:3420m_images}, to include all the directly arriving light. Then we calculate the mean gray value with 100 central pixels around the centroid from both \text{CamA} and \text{CamB} images as $I_{\text{A}}$ and $ I_{\text{B}}$ in formula \ref{I_center} and obtain the final attenuation length.

The uncertainty estimation for the $I_{\text{center}}$ method includes conventional systematic and statistical errors from the camera response, pre-calibration results of $I_{0}'/I_{0}$, the relative distances and the gray value processing. Systematic uncertainties such as luminosity loss at air-glass-water interface are canceled out in this setup.

To decode the absorption and scattering length from the attenuation process respectively, we apply the $\chi^2$ fitting method that compares the angular distribution of radiance between experimental data and the simulated data. 

Images are selected and cropped using the same procedure introduced in the $I_{\text{center}}$ method. Then both experimental data and simulated data of \text{CamA} and \text{CamB} are normalized with the same factor to keep the ratio of gray values of the center area unchanged. Considering the central symmetry, 2D images are converted into a 1D gray value array by calculating the mean gray value of those pixels which have the same pixel distance from the centroid. Finally, we choose the first 40 and 20 points in the gray value arrays of \text{CamA} and \text{CamB} to calculate the sum of $\chi^2$ pixel by pixel with the model:

\begin{equation}
    \chi^2 = \sum_{i=1}^{N}{ \frac{ [ M_i - T_i (1+\sum_{k=1}^{K}{\epsilon_{k}})]^2}{\sigma_{Mi}^2 + \sigma_{Ti}^2} } + \sum_{k=1}^{K}{ \frac{{\epsilon_{k}^2 }}{\sigma_{k}^2}} .
\end{equation}

Here, $M_i$ is the measured data in $i$th pixel bin while $T_i$ is the model prediction. $\sigma_{Mi}$ and $\sigma_{Ti}$ are both uncorrelated uncertainties coming from statistical fluctuation. $\epsilon$ is added to include correlated systematic uncertainties such as the uncertainties in distances and the calibration result of $I_{0}'/I_{0}$. Other uncertainties such as the slight de-focusing of the imaging process and the non-uniformity of the light emitter were all included in simulation and added as nuisance parameters when calculating $\chi^2$.

Due to the cameras' limited viewing angle, a numerical factor is required to account for the undetected scattered photons. This factor can be used to convert the canonical attenuation length to an effective attenuation length. This factor is calculated in the simulation assuming the best-fit optical parameters and uncertainties as input. 
The analysis results are summarized in Fig. \ref{fig:attenuation}.

\section*{Data availability}

All data supporting this study are provided in Results section and Extended Data of this paper. Extra data is available from the corresponding author upon reasonable request. A video of the small scale experiments performed in the ship towing tank is also provided from
the following URL: \url{https://trident.sjtu.edu.cn/en/gallery/videos/16/ship-towing-tank-experiments}

\section*{Code availability}
Code for optical process simulation in seawater can be found on \url{https://github.com/TRIDENT-Neutrino-Telescope/Pathfinder-Optical-Simulation}. Analysis and fitting for the optical properties were performed using ROOT and Python. Simulation and analysis for the detector performance were developed based on CORSIKA8, Geant4, OptiX and ROOT packages. Codes developed for data analysis are available from the corresponding author upon reasonable request.

\section*{Acknowledgements}

We are grateful to Frank Wilczek for his earnest support on jump starting this project. We are also appreciative for the helpful exchanges with Francis Halzen and Paschal Coyle on a number of issues involving stable operation of neutrino telescopes. Moreover, the authors thank Yifang Wang for the insightful conversations which have helped greatly in shaping of this project. The sea scouting team owes special gratitude to Shuangnan Zhang for his company and support during the voyage. We are obliged to Jie Zhang and Zhongqin Lin – the former and current president of Shanghai Jiao Tong University (SJTU), respectively – for their enthusiasm and enduring support for this proposal. Last but not least, we are much indepted to Dan Wu – the former vice president of SJTU — who helped assemble and coordinate the interdisplinary service team, without which this pathfinder experiment would not have been possible.

We thank the China Jinping Underground Laboratory for providing precise measurements of the seawater radioactivity. We are also grateful for the 1:2-million maps of the South China Sea and relevant data provided by the Guangzhou Marine Geological Survey, Ministry of Natural Resources of China. D. L. Xu is sponsored by the Ministry of Science and Technology of China under grant No. 2022YFA1605500, Office of Science and Technology, Shanghai Municipal Government under grant No. 22JC1410100, the National Natural Science Foundation of China (NSFC) under grant No. 12175137, and also by SJTU under the Double First Class startup fund and the Foresight grants No. 21X010202013 and No. 21X010200816; X. L. Tian is supported by the NSFC grant No. U20A20328; W. H. Wu is supported by the Shanghai Pujiang Program under grant No. 20PJ1409300; and J. L. Liu is under the sponsorship from the Hongwen Foundation in Hong Kong and Tencent Foundation in China.



\section*{Author contributions}

Z. P. Ye, F. Hu and W. Tian contributed substantially and equally throughout the apparatus development and data analyses; I. Morton-Blake led the sensitivity analyses and manuscript review; F. Y. D. Zhang made significant contributions to the manuscript editing; For the pathfinder experiment, W. H. Wu led the development of the electronics and DAQ systems, G. H. Gong developed the clock synchronization system, X. L. Tian designed the deployment device and led the marine operations; D. L. Xu proposed the TRIDENT project, acting as the collaboration's spokesperson and led the pathfinder experiment. All authors have read and consent on the content. The authors of this work are all members of the TRIDENT collaboration.

\section*{Competing Interests}
The authors declare no competing interests.

\section*{Additional Information}

Correspondence and requests for materials related to this work should be addressed to Donglian Xu (ORCID: \url{https://orcid.org/0000-0003-1639-8829}).

\nocite{Herold_40K_simulation:2017}
\nocite{Wiebusch:2003}
\nocite{Pythia8:2014, Cooper-Sarkar:2011}
\nocite{PROPOSAL:2013}
\nocite{CORSIKA8:2018}
\nocite{Opticks:2019}


\clearpage


\clearpage

\clearpage
\renewcommand\thefigure{\arabic{figure}}
\setcounter{figure}{0}
\renewcommand\thetable{\arabic{table}}
\setcounter{table}{0}

\clearpage
\begin{figure}[htbp]
  \renewcommand\figurename{\textbf{Extended Data Fig.}}
    \centering  
    \figuretitle{Geographical information of the TRIDENT site.}
    \includegraphics[width=0.85\linewidth]{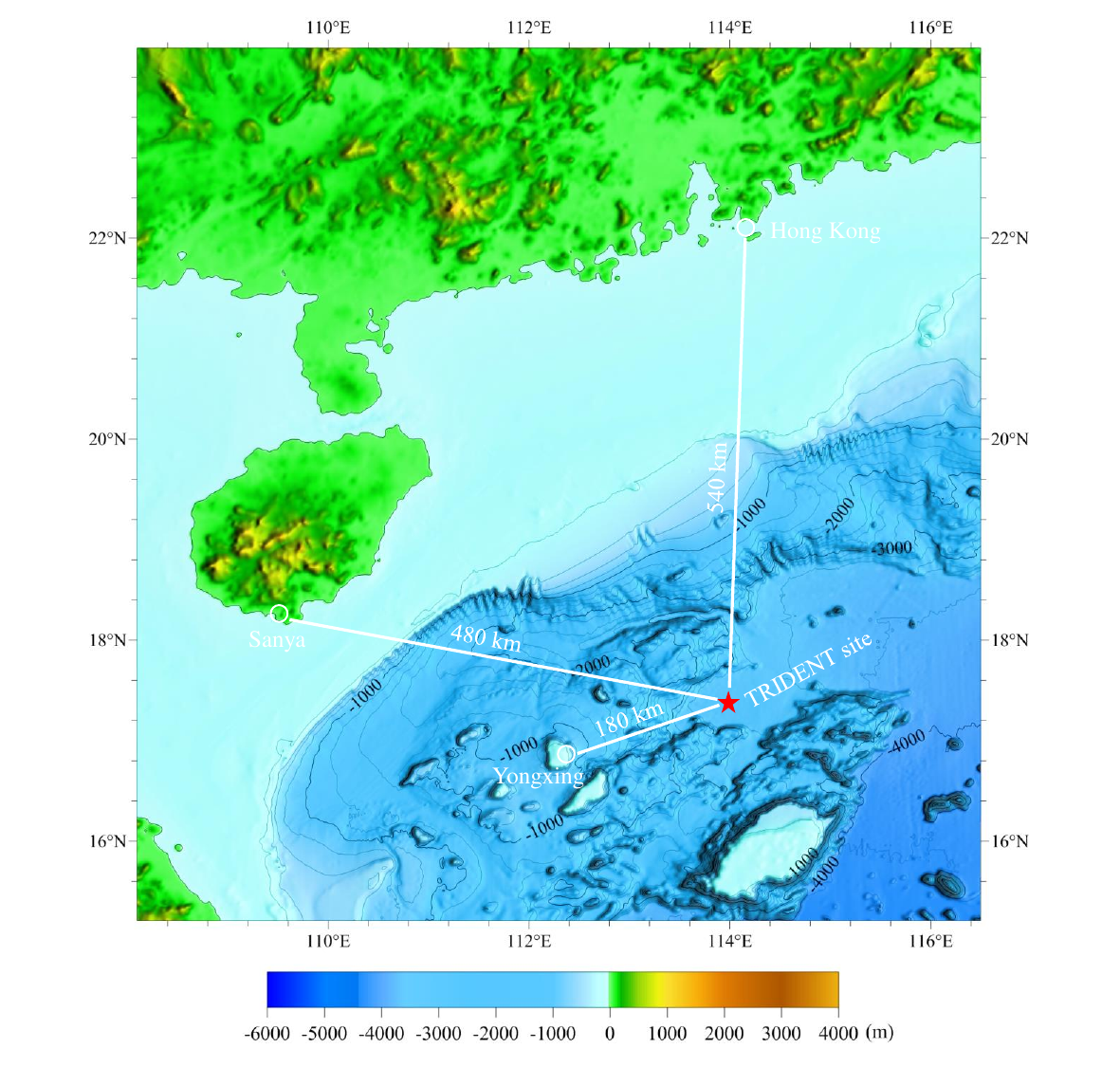}
    \caption{The selected site is marked by the red star in this map \cite{site_info}. The distance between the TRIDENT site and nearby cities are shown by the white lines. The nearest island with infrastructure, Yongxing Island, is $180~\mathrm{km}$ away.} 
    \label{fig:geographic_map}
\end{figure} 

\clearpage
\begin{figure}[!htb] 
    \renewcommand\figurename{\textbf{Extended Data Fig.}}
    \begin{subfigure}[!htb]{0.90\textwidth}
    \centering
    \figuretitle{Oceanographic conditions of the TRIDENT site.}
    \includegraphics[width=0.42\linewidth]{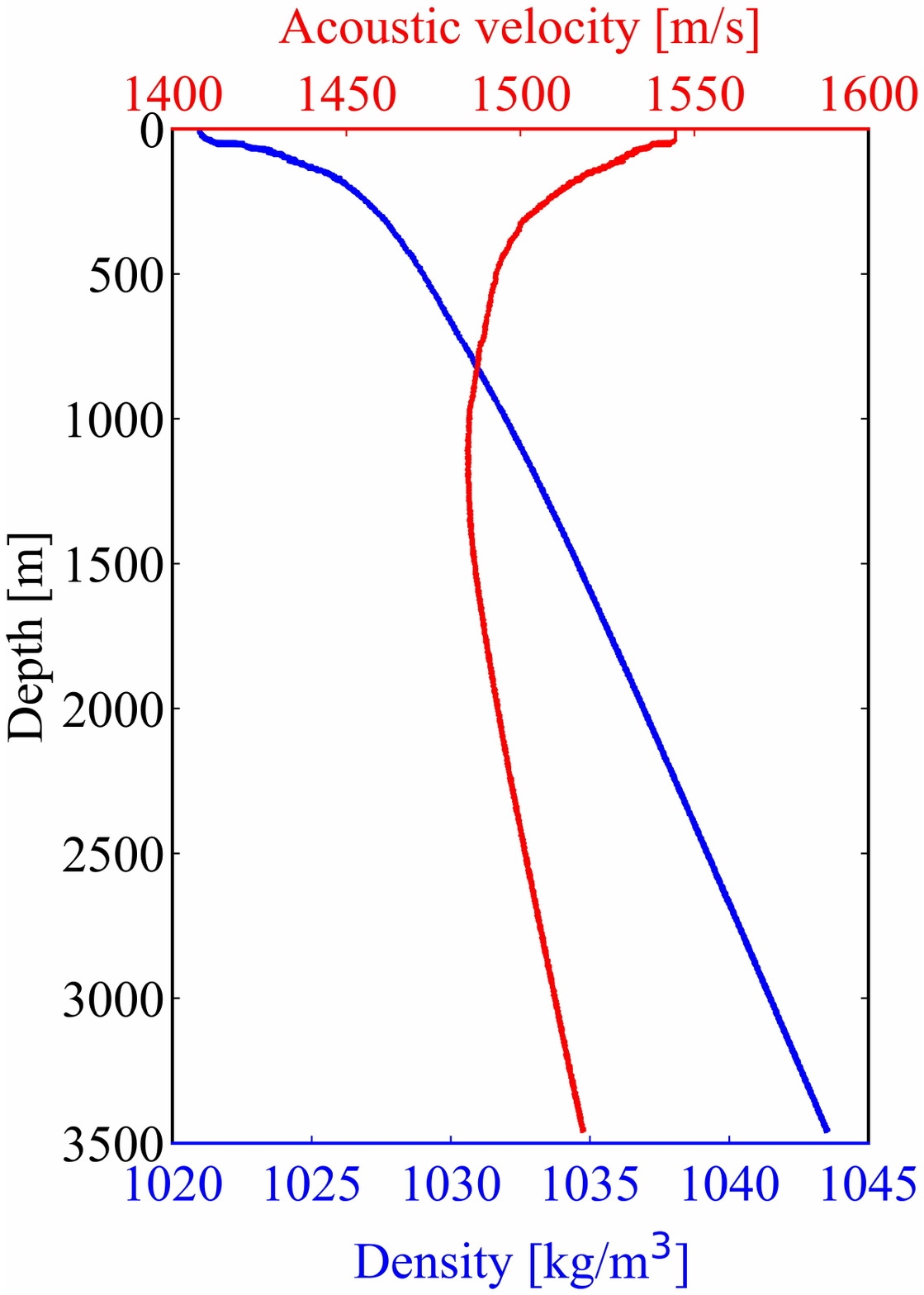}
    \includegraphics[width=0.42\linewidth]{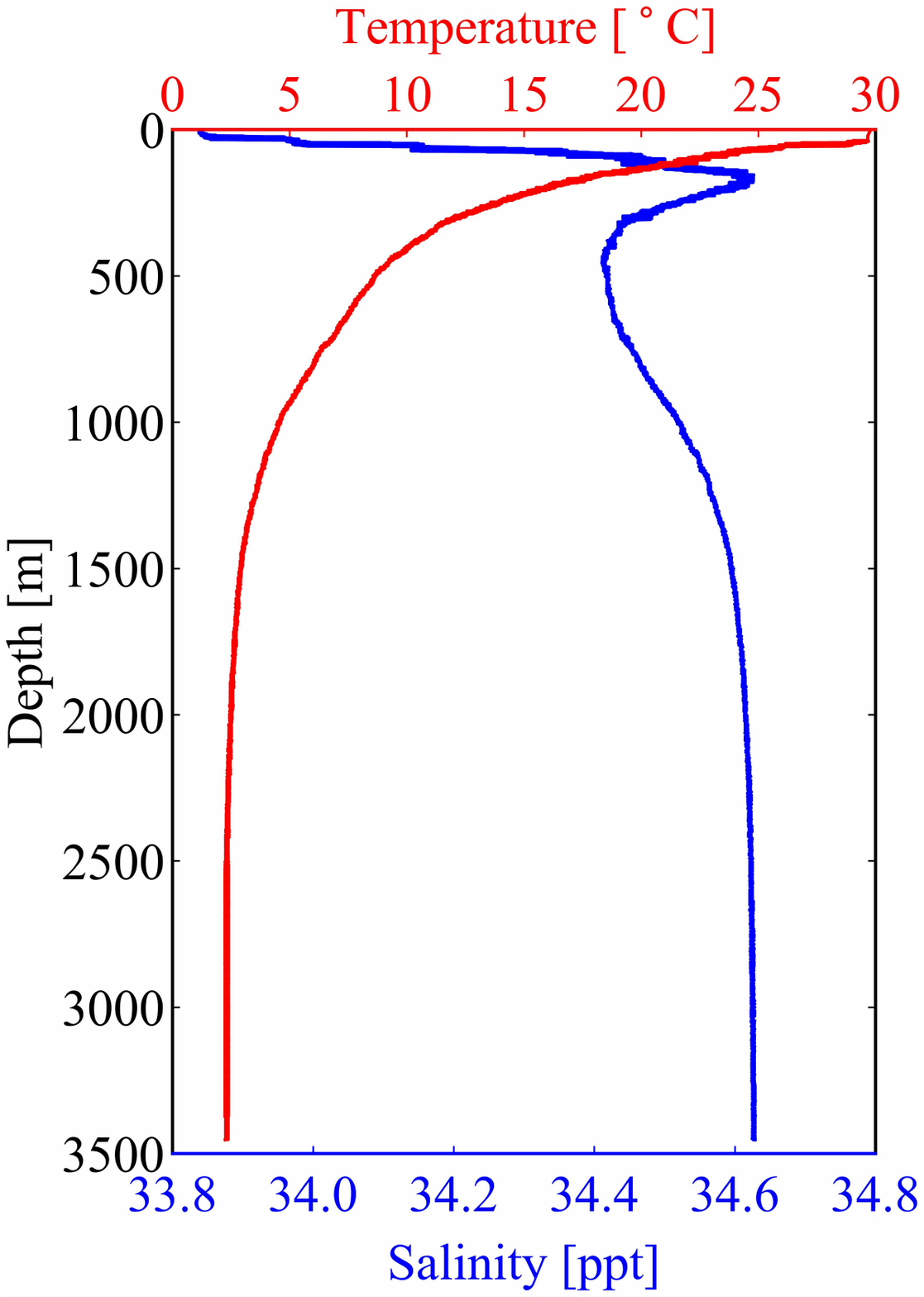}
    \caption{Left panel: water density and sound speed as a function of depth. Right panel: temperature and salinity as a function of depth.} 
    \end{subfigure}
    \begin{subfigure}[!htb]{0.90\textwidth}
    \centering
    \includegraphics[width=0.95\linewidth]{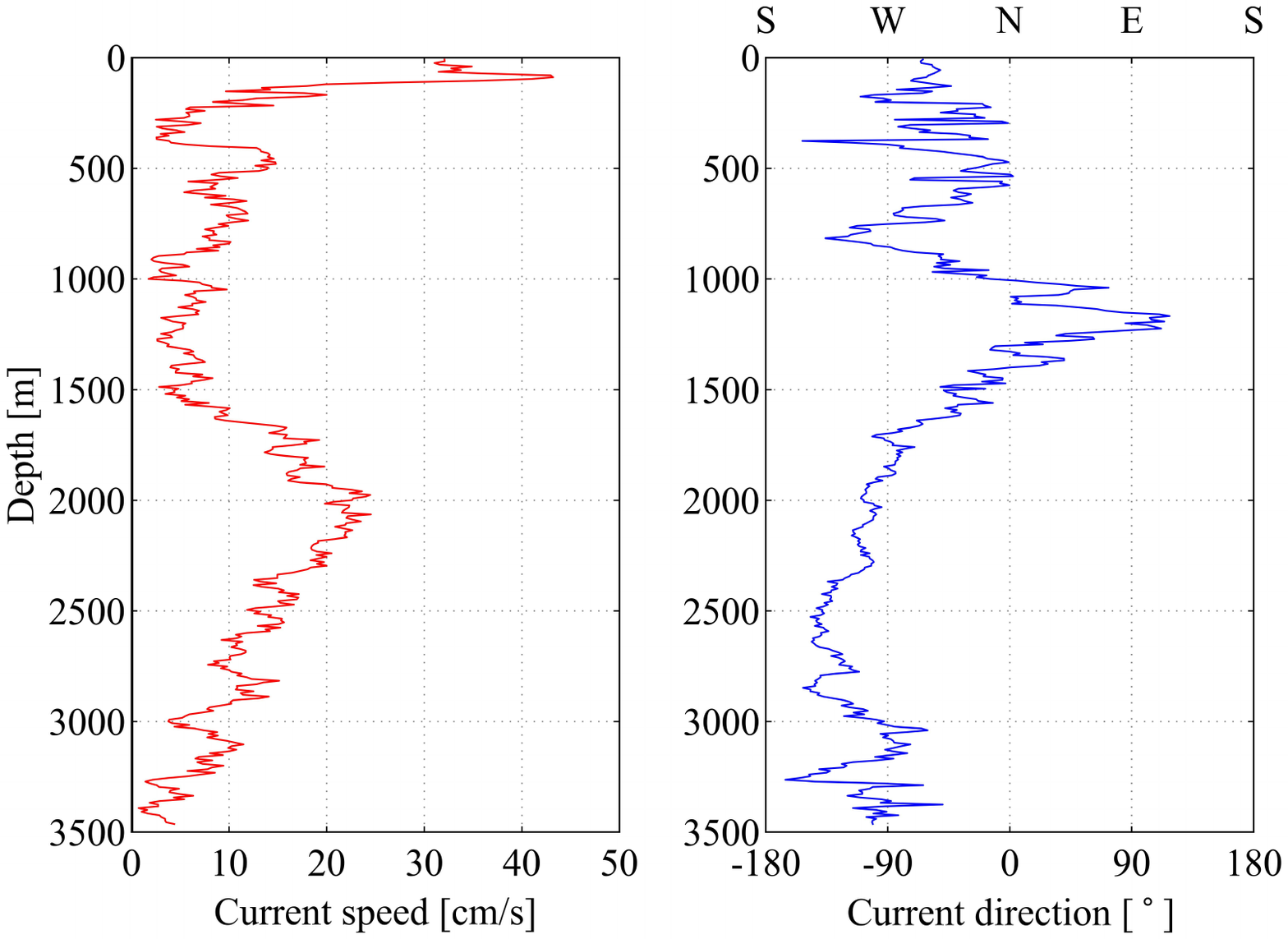}
    \caption{Current speed and direction as a function of depth.} 
    \end{subfigure}
    \caption{The water density, sound speed, temperature, salinity, water current speed and direction were measured as a function of depth at the TRIDENT site. The data were taken on September 6, 2021.}
    \label{fig:ocean_condition}
\end{figure}

\clearpage
\begin{figure}[htbp]%
  \renewcommand\figurename{\textbf{Extended Data Fig.}}
    \centering
    \figuretitle{Expected hit rate per 3-inch PMT caused by Potassium-40 radioactivity.}
    \includegraphics[width=0.85\textwidth]{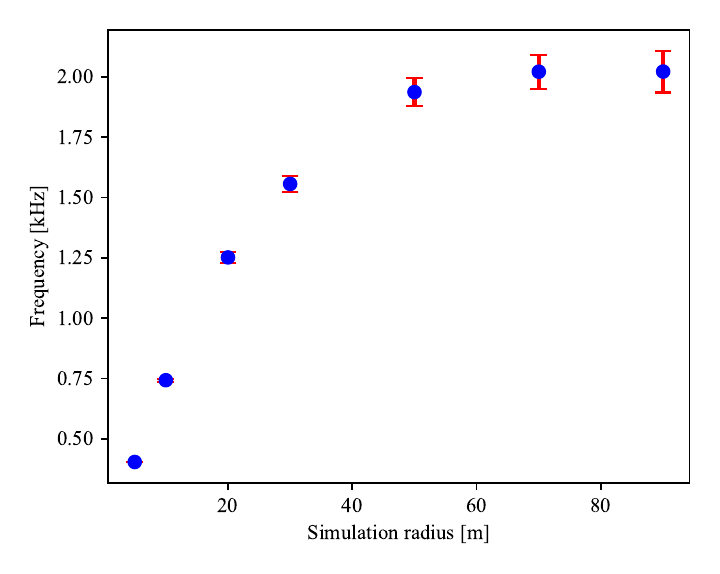}
    \caption{The hit rate converges to $\sim$ 2 kHz as the implemented spherical volume increases. The simulation setup is adapted from \cite{Herold_40K_simulation:2017}, in which the detector layout is the same as the light receiver module used in T-REX, and $^{40}$K decay vertices are sampled uniformly within a spherical volume whose center is the detector and radius is a variable. The $^{40}$K abundance is set to be 10.78 Bq/kg and their decay will produce Cherenkov photons that can hit the PMTs. For each data point, $3 \times 10^9$ $^{40}$K decay events are simulated in \textsf{Geant4}. Error bars reflect poisson statistical uncertainties in simulation.}
    \label{fig:k40_simulation}
\end{figure}

\clearpage  
\begin{figure}[htbp]
  \renewcommand\figurename{\textbf{Extended Data Fig.}}
    \centering
    \figuretitle{A schematic plot of the T-REX apparatus.}
    \includegraphics[width=0.85\linewidth]{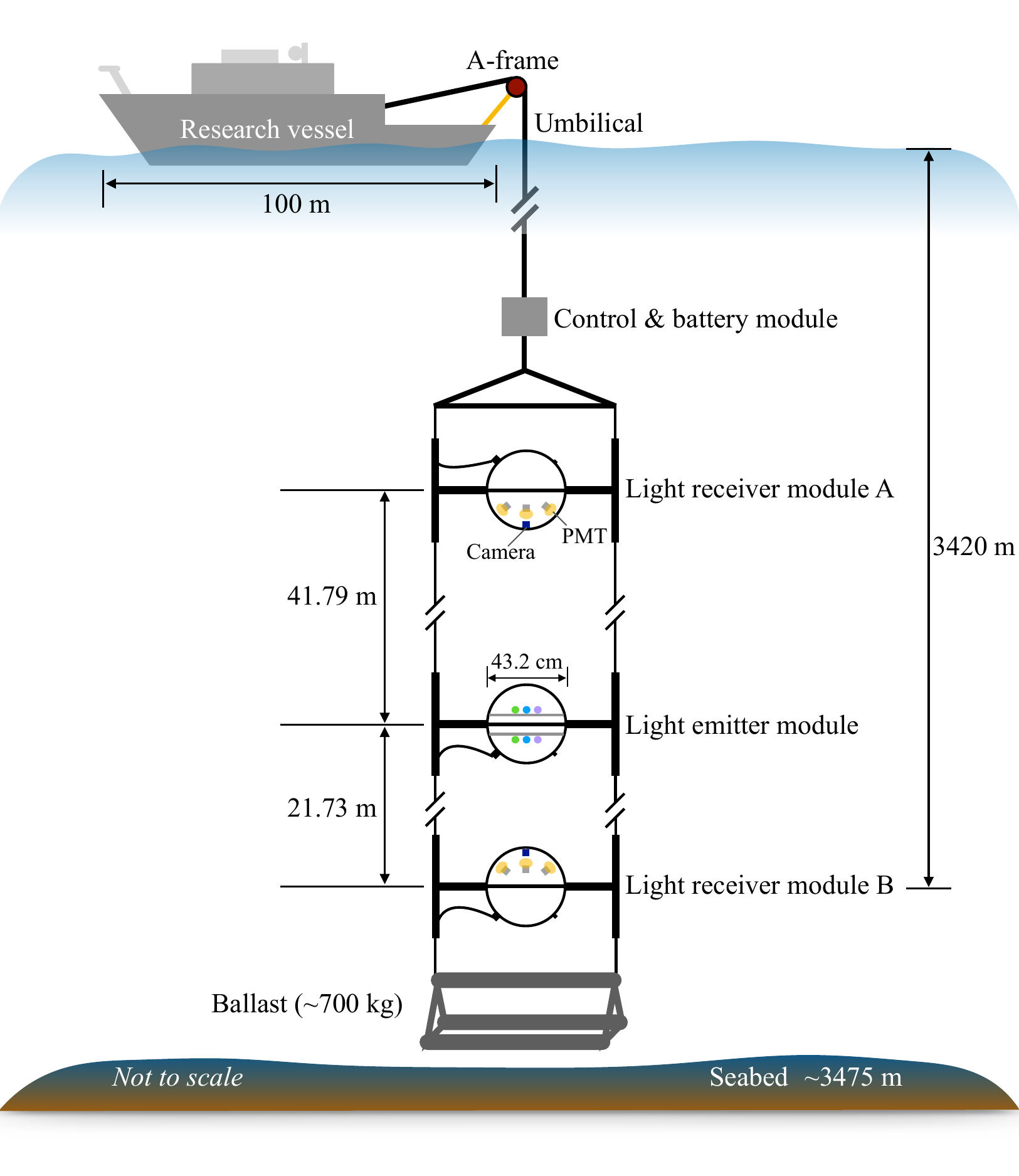}
    \caption{The apparatus was deployed to a depth of $3420~\mathrm{m}$ by the umbilical cable from a research vessel. A light emitter module is placed in the middle. Light receiver module A and B are placed on the top and bottom at distances of $41.79\pm 0.04 ~\mathrm{m}$ and $21.73\pm 0.02 ~\mathrm{m}$ respectively. Each module is connected to the control and battery module via a composite optical-electric cable for data communications and power supply. This figure is not to scale. }
    \label{fig:trex_apparatus}
\end{figure} 

\clearpage
\begin{figure}[htbp]%
  \renewcommand\figurename{\textbf{Extended Data Fig.}}
    \centering
    \figuretitle{
    Measured and best-fit PMT photon arrival time distributions.}
    \includegraphics[width=0.9\linewidth]{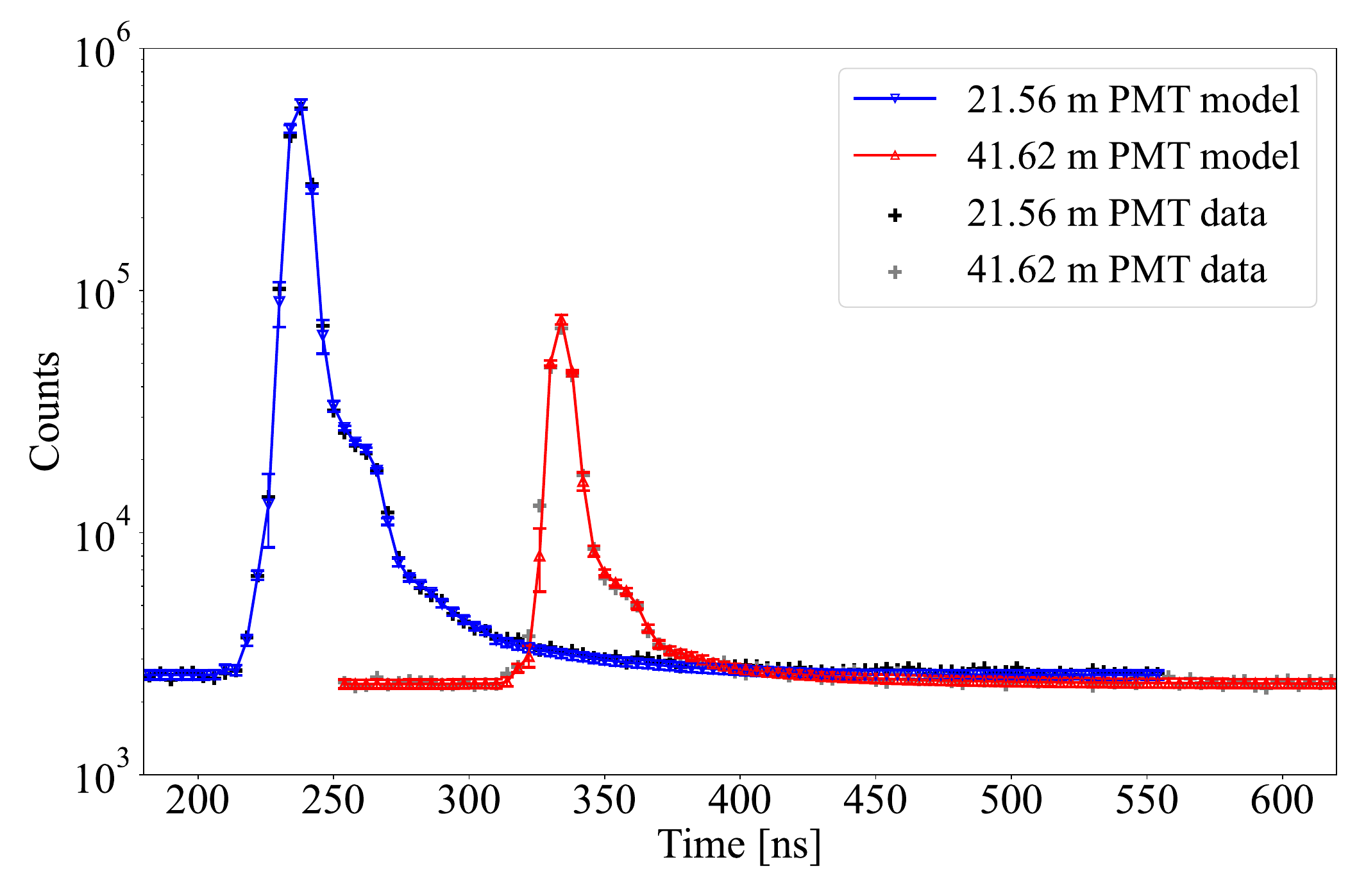}
    \caption{
    Data shown are measured at the wavelength of $450~\mathrm{nm}$. Points and error bars in the fitting model represents the best-fit results and the 68\% confidence-level regions in the $\chi^2$ fitting. Data sample used for the analysis contains $\sim 5 \times 10^6$ and $\sim 7 \times 10^5$ detected photons for a pair of PMTs at 21.56 m and 41.62 m, respectively.
    The intrinsic width of the LED light pulses is $\sim 3~\mathrm{ns}$, however, some photons are scattered and arrive later. The tail of the distribution is thus caused by scattering effect. The ratio of the number of photons detected by the top and bottom PMTs are determined by the distance ratio and absorption strength.
    The photon arrival time model is obtained by \textsf{Geant4} simulation, with calibrated LED and PMT response taken into account. 
    A $\chi^2$ fitting method is used to evaluate the goodness of fit.     Minimization of the $\chi^2$ returns the best-fit model, obtaining the measurement results for optical parameters, including absorption length, Rayleigh scattering length, Mie scattering length, Mie scattering mean angle, and refraction index.} 
    \label{fig:pmt_fitting}
\end{figure}

\clearpage
\begin{figure}[htbp]
 \renewcommand\figurename{\textbf{Extended Data Fig.}}
\centering
\figuretitle{Example images of the light emitter recorded by the camera system at the depth of 3420 m.}
\includegraphics[width=.92\textwidth]{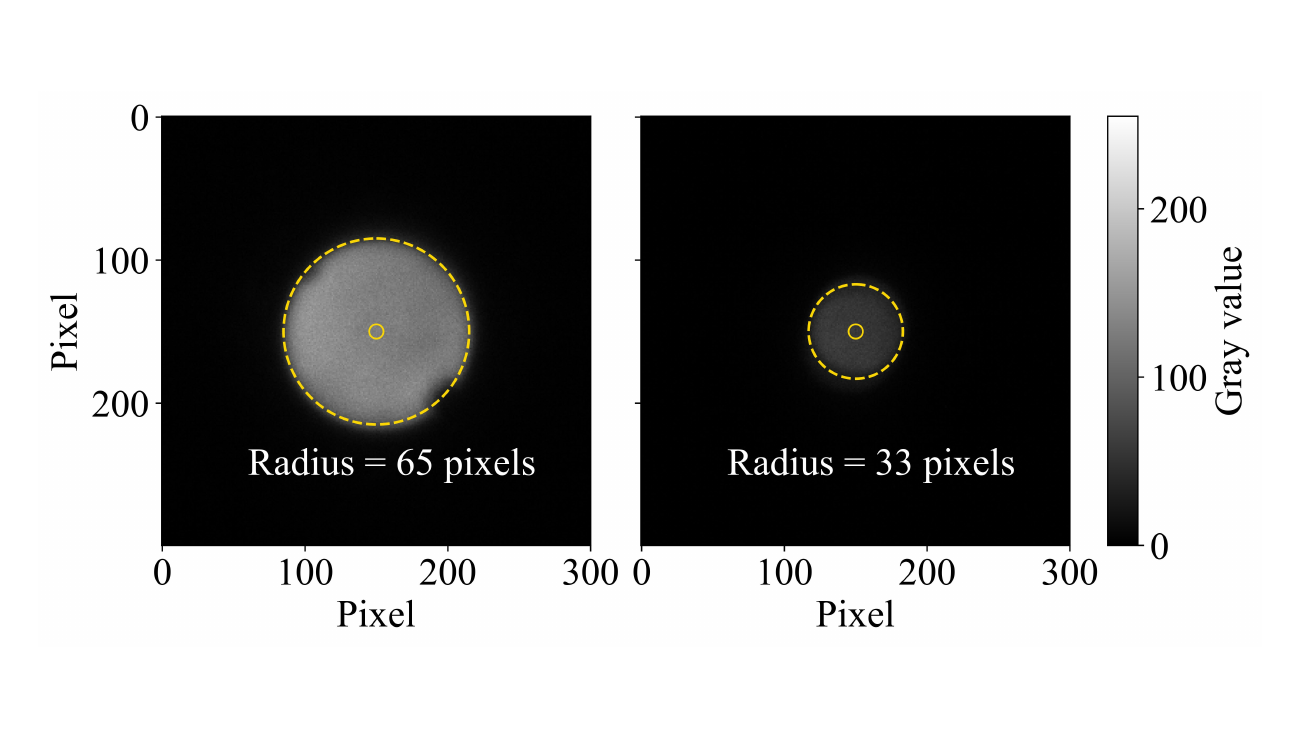}
\caption{Images of the light emitter at a wavelength of $460~\mathrm{nm}$ with an exposure time of $0.02~\mathrm{s}$ at a water depth of $3420~\mathrm{m}$. The dashed circle shows the profile of the light emitter and the gray value distribution of the inside region is used for the $\chi ^2$ fitting analysis. The region enclosed by the solid circle is used for the analysis of the $I_\mathrm{center}$ method. Left panel: an image taken by CamB at a distance of $21.5~\mathrm{m}$, the radius of the dashed circle is 65 pixels.  Right panel: an image taken by CamA at a distance of $41.6~\mathrm{m}$, the radius of the dashed circle is 33 pixels.}
\label{fig:3420m_images}
\end{figure}

\clearpage
\begin{figure}[htbp]
\renewcommand\figurename{\textbf{Extended Data Fig.}}
\centering
\figuretitle{Angular resolution of TRIDENT for muon neutrino charged-current events.}
\includegraphics[width=0.9\textwidth]{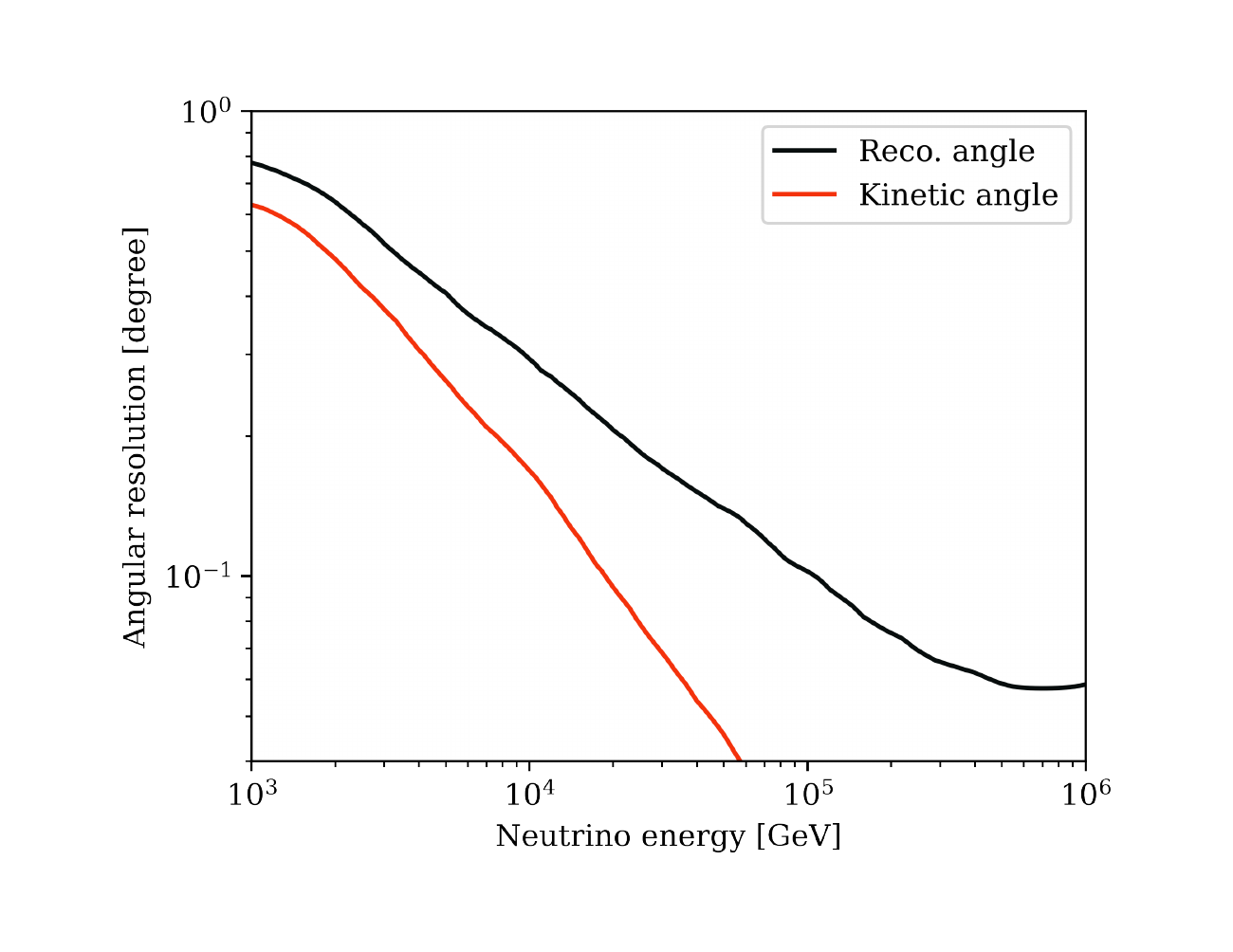}
\caption{The black line represents the open angle between the reconstructed muon track direction and the primary neutrino direction. The red line is the open angle between the primary neutrino direction and the true muon direction. At an energy of $\sim100~\mathrm{TeV}$, the angular error is expected to reach 0.1$^{\circ}$. A multi-photon-electron algorithm \cite{Wiebusch:2003} that utilizes both fast time resolution of SiPMs and large photon collection area of PMTs \cite{Hu:2021jjt} is used for reconstruction.}
\label{fig:angular_resolution} 
\end{figure} 

\clearpage
\begin{figure}[htbp]
\renewcommand\figurename{\textbf{Extended Data Fig.}}
\centering
\figuretitle{Effective areas of TRIDENT for muon neutrino charged-current events.}
\includegraphics[width=0.72\textwidth]{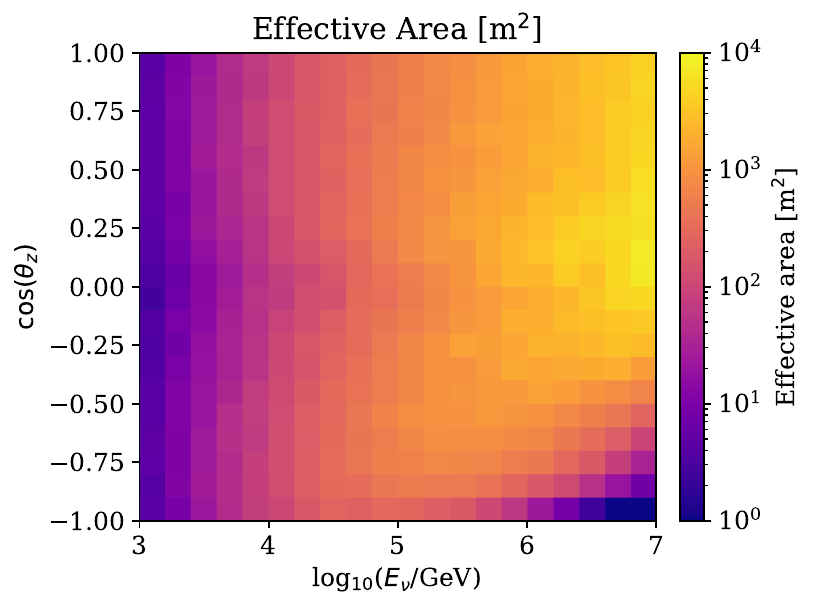}
\includegraphics[width=0.62\textwidth]{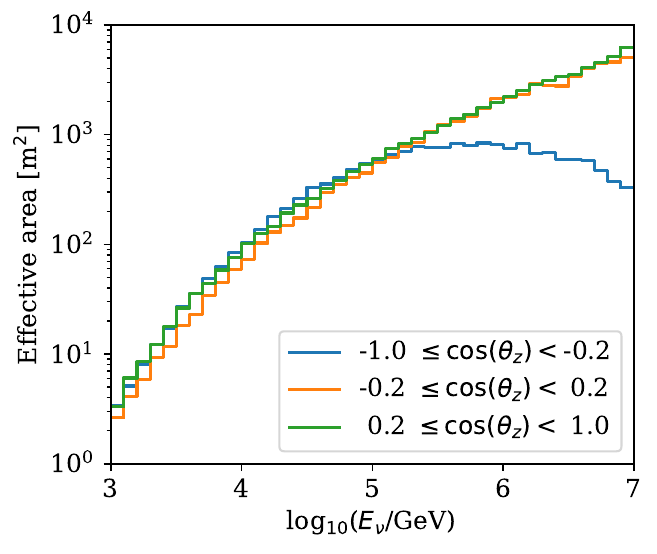}
\caption{Top panel: effective area shown as a function of primary neutrino energy and zenith angle in TRIDENT. Bottom panel: effective area shown as a function of primary neutrino energy at three zenith bands. 
At an energy of $\sim100~\mathrm{TeV}$, the effective area for up-going events is expected to reach $5 \times 10^2 ~\mathrm{m^2}$. Only events with reconstructed angular error less than 6 degrees are used in these effective area calculations.
The neutrino-nucleon CC interaction vertex is generated around detector via \textsf{Pythia8} \cite{Pythia8:2014, Cooper-Sarkar:2011}. The muons are propagated to the detector by \textsf{PROPOSAL} \cite{PROPOSAL:2013}, using \textsf{CORSIKA8} simulation framework \cite{CORSIKA8:2018}. For particles arriving to the detector, \textsf{Geant4} simulation is used for particle simulation and \textsf{OptiX} for Cherenkov photon generation and propagation \cite{Opticks:2019}.}
\label{fig:effective_area} 
\end{figure}

\clearpage
\begin{table}[ht!]
\renewcommand\tablename{\textbf{Extended Data Table}}
\begin{subtable}[t]{\textwidth}
\caption{Optical parameters measured by the PMT system at 405 nm at the depth of 3420 m.}
\begin{tabular*}{\textwidth}{@{\extracolsep{\fill}}lrrrrr}
\toprule
Method & $\lambda_{\text{abs}}$ [m] & $\lambda_{\text{ray}}$ [m] & $\lambda_{\text{mie}} $[m] & $\cos \theta_{\text{mie}}$ & $\lambda_{\text{att}}$ [m] \\
\midrule
$\chi^2$ fitting & $ 20.8^{+1.0}_{-0.9} $ & $ 120^{+12}_{-9} $ & $ 55^{+20}_{-10}$ & $ 0.98 ^{+ 0.02}_{-0.02} $ & $13.4^{+5.1}_{-2.7}$ \\ 
MCMC & $19.2^{+0.5}_{-0.5}$ & $114^{+4}_{-3}$ & $46^{+6}_{-15}$ & $0.98^{+0.01}_{-0.01}$ & $12.1^{+0.4}_{-1.2}$ \\
\bottomrule
\end{tabular*}
\end{subtable}
\vspace*{5mm}
\begin{subtable}[t]{\textwidth}
  \caption{Optical parameters measured by the PMT system at 525 nm at the depth of 3420 m.}
\begin{tabular*}{\textwidth}{@{\extracolsep{\fill}}lrrrrr}
\toprule
Method & $\lambda_{\text{abs}}$ [m] & $\lambda_{\text{ray}}$ [m] & $\lambda_{\text{mie}} $[m] & $\cos \theta_{\text{mie}}$ & $\lambda_{\text{att}}$ [m] \\
\midrule
$\chi^2$ fitting & $ 22.2^{+1.2}_{-1.0}$ & $ 330^{+20}_{-15} $ & $ 135^{+30}_{-20}$ & $0.98^{+0.02}_{-0.02}$ & $18.0^{+4.3}_{-2.9}$ \\ 
MCMC & $18.5^{+0.7}_{-0.6}$ & $330^{+19}_{-22}$ & $127^{+13}_{-22}$ & $0.98^{+0.01}_{-0.01}$ & $15.3^{+0.5}_{-0.5}$\\
\bottomrule
\end{tabular*}
\end{subtable}
\caption{Results of optical property measurements from the PMT system at wavelengths of $405~\mathrm{nm}$ and $525~\mathrm{nm}$ and at the depth of $3420~\mathrm{m}$. The data collection for each wavelength lasted $\sim 10$ minutes. Error bars reflect both statistical and systematic uncertainties.}
\label{tab:pmt_extend_results}
\end{table}

\clearpage
\begin{table}[ht!]
\renewcommand\tablename{\textbf{Extended Data Table}}
\begin{subtable}[t]{\textwidth}
\caption{Optical parameters measured by the camera system with $I_{\text{center}}$ method at different wavelengths and depths.}
\begin{tabular*}{\textwidth}{@{\extracolsep{\fill}}ccc}
\toprule
Wavelengths [nm] & Depth [m] & $\lambda_{\text{att}}$ [m] \\
\midrule
460 & 1221 & $17.8 \pm 1.1 $ \\ 
460 & 2042 & $18.7 \pm 1.2 $  \\
460 & 3420 & $19.3 \pm 1.3 $  \\
525 & 3420 & $14.6 \pm 0.7$ \\
405 & 3420 & $13.7 \pm 0.6$ \\
\bottomrule
\end{tabular*}
\end{subtable}
\vspace*{5mm}
\begin{subtable}[t]{\textwidth}
\caption{Optical parameters measured by the camera system with $\chi^2$ fitting at 525 nm.}
\begin{tabular*}{\textwidth}{@{\extracolsep{\fill}}ccccc}
\toprule
Wavelengths [nm] & Depth [m] & $\lambda_{\text{abs}}$ & $\lambda_{\text{sca}}$ [m] & $\lambda_{\text{att}}$ [m] \\
\midrule
525 & 3420 & $17.5\pm0.5$ & $57.2\pm6.5$ & $13.4\pm0.2$ \\
\bottomrule
\end{tabular*}
\end{subtable}
\caption{Results of attenuation lengths at $460~\mathrm{nm}$ in different depths and other optical parameters at $525~\mathrm{nm}$ and $405~\mathrm{nm}$ at the depth of $3420~\mathrm{m}$ measured by the camera system. The data collection at each wavelength and each depth lasted $\sim 8$ minutes. }
\label{tab:camera_extend_results}
\end{table}

\end{document}